\documentclass[aps,prd,reprint,twocolumn,10pt,eqsecnum,showpacs,nofootinbib,amsfonts,amssymb,amsmath,longbibliography,superscriptaddress,floatfix]{revtex4-1}
\usepackage[colorlinks=true,allcolors=blue, pdfusetitle]{hyperref}
\usepackage{bm}
\usepackage{mathrsfs}
\usepackage[dvipsnames]{xcolor}
\usepackage{mathtools}
\usepackage{graphicx}
\usepackage{orcidlink}
\usepackage{tensor}

\usepackage{algorithm}
\usepackage{algpseudocode}

\usepackage[commentmarkup=uwave]{changes}
\definechangesauthor[name=hs, color=cyan]{hs}

\usepackage{fontawesome5}

\usepackage{ragged2e}

\usepackage{tikz}

\usetikzlibrary{shapes.geometric, arrows,positioning}

\tikzstyle{startstop} = [rectangle, rounded corners, minimum width=3cm, minimum height=1cm, text centered, draw=black, fill=orange!30]
\tikzstyle{process} = [rectangle,rounded corners, minimum width=3cm, minimum height=1cm, text centered, draw=black, fill=blue!15]
\tikzstyle{decision} = [rectangle,rounded corners, minimum width=3cm, minimum height=1cm, text centered, draw=black, fill=green!20]
\tikzstyle{arrow} = [->, >=stealth]

\graphicspath{{../plots/}}

\newcommand{\algQNMFinder}{%
	\begin{figure*}
		\begin{minipage}{\linewidth}
			\begin{algorithm}[H]
				\caption{\textbf{\emph{qnmfinder}}}
				\label{alg:qnmfinder}
				\begin{algorithmic}[1]
					\Function{build\_model($h$, $t_{0}^{i}$, $t_{0}^{f}$, $\Delta t_{0}$, $\Delta\omega^{\max}$, $\Delta A^{\min}$, $\sigma^{\max}$, $N_{\mathrm{free}}^{\max}$)}{} : 
					\State $t_0=t_0^f$; $N_{\mathrm{free}}=1$; model = \Call{Model}{ };
					\While {$t_0>t_{0}^{i}$}: 
					\State $(\ell,m)$ =  \Call{most\_unmodeled\_mode}{}($h$, model)
					\State qnm = \Call{best\_fitting\_qnm}{}($h$, model, $(\ell,m)$, $\Delta\omega^{\max}$, $N_{\mathrm{free}}$)
					\If{qnm \textbf{is} \Call{None}{}}
					\State $t_{0}=t_{0}-\Delta t_{0}$
					\State\Call{continue}{}
					\EndIf
					\State fit\_model = \Call{linear\_least\_squares\_fit}{}($h$, model + qnm, \Call{arange}{$t_{0}^{i}$, $t_{0}^{f}$, $\Delta t_{0}$})
					\If{\Call{is\_stable}{}(fit\_model, $\Delta A^{\min}$, $\sigma^{\max}$)}
					\State model = model + qnm
					\State $t_{0}=t_{0}^{f}$
					\Else
					\If{$N_{\mathrm{free}}<N_{\mathrm{free}}^{\max}$}
					\State $N_{\mathrm{free}}=N_{\mathrm{free}}+1$
					\Else
					\State $t_{0}=t_{0}-\Delta t_{0}$
					\EndIf
					\EndIf
					\EndWhile
					\State\Return model
					\EndFunction
					\Function{most\_unmodeled\_mode}{}($h$, model) : 
					\State\Return $h.\mathrm{LM}$\Call{[argmax}{$\int_{t_{0}}^{t_{f}}\int_{S^{2}}\phantom{}_{-2}Y_{(\ell,m)}^{*}\left|\frac{d}{dt}\left(h - \mathrm{model}\right)\right|^{2}\,d\Omega\,dt$}\Call{]}{}
					\EndFunction
					\Function{best\_fitting\_qnm}{}($h$, model, $(\ell,m)$, $\Delta\omega^{\max}$, $N_{\mathrm{free}}$) : 
					\State $\omega$ = \Call{varpro}{}($h$, model, $(\ell,m)$, $N_{\mathrm{free}}$)
					\State $\Delta\omega$ = \Call{[}{}\Call{abs}{$\omega - \mathrm{qnm}.\omega$} \textbf{for} qnm \textbf{in} qnms\Call{[where}{qnms.$m==m$}\Call{]}{}\Call{]}{}
					\If{\Call{min}{$\Delta\omega$} $<\Delta\omega^{\max}$}
					\State\Return qnms\Call{[where}{qnms.$m==m$}\Call{]}{}\Call{[argmin}{$\Delta\omega$}\Call{]}{}
					\EndIf
					\EndFunction
					\Function{is\_stable}{}(fit\_model, $\Delta A^{\min}$, $\sigma^{\max}$) : 
					\For{qnm \textbf{in} fit\_model.qnms}
					\State$t_{0}^{1}=t_{0}^{i}$
					\State $\Delta t_{\mathrm{QNM}}=-\ln(\Delta A^{\min})/$\Call{Im}{qnm.$\omega$}
					\State $\sigma=\inf$
					\While{$t_{0}^{1}+\Delta t_{\mathrm{QNM}}<t_{0}^{f}$}
					\State $\sigma$ = \Call{min}{}($\sigma$, $\sqrt{\Delta\mathrm{Re}[\mathrm{qnm}.C[t_{0}^{1}:t_{0}^{1}+\Delta t_{\mathrm{QNM}}]]^2+\Delta\mathrm{Im}[\mathrm{qnm}.C[t_{0}^{1}:t_{0}^{1}+\Delta t_{\mathrm{QNM}}]]^2}/|\mathrm{qnm}.C|$)
					\State $t_{0}^{1}=t_{0}^{1}+\Delta t_{0}$
					\EndWhile
					\If{$\sigma>\sigma^{\max}$}
					\State\Return\Call{False}{}
					\EndIf
					\EndFor
					\State\Return\Call{True}{}
					\EndFunction
				\end{algorithmic}
			\end{algorithm}
	\end{minipage}
\RaggedRight{$h$ is the NR waveform, $t_{0}^{i}$ is the earliest fitting start time, $t_{0}^{f}$ is the latest fitting start time, $\Delta t_{0}$ is the amount by which $t_{0}$ is changed after each failed iteration, $\Delta\omega^{\mathrm{max}}$ is the norm tolerance for a QNM to be considered a match to a \texttt{VarPro} frequency, $\Delta A^{\mathrm{min}}$ is the same as in Eq.~\eqref{eq:timerequirement}, $\sigma^{\mathrm{max}}$ is the same as in Eq.~\eqref{eq:variationrequirement}, and $N_{\mathrm{free}}^{\mathrm{max}}$ is the maximum number of free frequencies. Note that in Line 40, $\Delta$ represents the standard deviation of the real or imaginary part of the QNM amplitude.}
\end{figure*}
}

\newcommand{\figFlowChart}{%
	\begin{figure*}[tb]
			\begin{tikzpicture}
				\node (start) [startstop,xshift = 0.0] {Is $t_{0} > t_{0}^{i}$?};
				\node(continue)[process,below of =start, yshift = -1cm,xshift = 0.4cm]{
					\begin{tabular}{c}
						Find most unmodeled\\$(\ell,m)$ mode
					\end{tabular}
				};       
				\node(return)[process,above of =start, yshift = 0.5cm, xshift = 0.4cm, fill=green!20]{Return model};     
				\node (process1) [process,right of=continue, xshift = 2.6cm, fill=orange!30] 
				{
					\begin{tabular}{c}
						$N_\mathrm{free}$ QNMs\\match \texttt{VarPro}?
					\end{tabular}
				};      
				\node (process2) [process, below of=process1, yshift = -0.3cm, xshift = 3.5cm, fill=orange!30] {
					\begin{tabular}{c}
						Are the QNMs stable?
					\end{tabular}
				};
				\node (process3) [process, below of=process1,yshift = 2.2cm,xshift = 3.5 cm, fill = red!20] {$t_{0} \to t_{0} - \Delta t_{0}$};
				\node (process4) [process, right of=process2, yshift=2.34cm,xshift = 2.45cm, fill = green!10] {
					\begin{tabular}{c}
						Add QNMs \\ to the model; \\
						$t_{0} \to t_{0}^{f}$
					\end{tabular}
				};
				\node (process5) [process, right of=process2, yshift=-1.2cm,xshift = 2.45cm,fill=orange!30] {Is $N_\mathrm{free} < N_\mathrm{max}$?};  
				\node (process6) [process, right of=process5, yshift=3.7cm,xshift = 2.4cm,fill = red!20] {$t_{0} \to t_{0} - \Delta t_{0}$};       
				\node (process7) [process, right of=process5, yshift=-1.2cm,xshift = 2.4cm,fill = red!20] {$N_\mathrm{free} \to N_\mathrm{free} + 1$};

				\draw [arrow] (start.south) -- node[anchor=east,xshift=30pt] {Yes} (continue.north);
				\draw [arrow] (start.north) -- node[anchor=west,xshift=10pt] {No} (return.south);
				\draw [arrow] (continue) -- (process1);
				\draw [arrow] (process1.east) -- node[anchor=east,xshift=30pt] {Yes} (process2.north);
				\draw [arrow] (process1.east) -- node[anchor=west,xshift=10pt] {No} (process3.south);
				\draw [arrow] (process2.east) -- node[anchor=east,xshift=30pt] {Yes} (process4.south);
				\draw [arrow] (process2.east) -- node[anchor=west,xshift=10pt] {No} (process5.north);
				\draw [arrow] (process5.east) -- node[anchor=east,xshift=25pt] {No} (process6.south);
				\draw [arrow] (process5.east) -- node[anchor=west,xshift=8pt] {Yes} (process7.north);
				\draw [arrow] (process3.north) |- (start.east);
				\draw [arrow] (process4.north) |- (start.east);
				\draw [arrow] (process6.north)  |- (start.east);
				\draw [arrow] (process7.east) -- ++(0.2,0) node[midway, above] {} |- (start.east);
			\end{tikzpicture}    
	\caption{
		Pictorial version of Alg.~\ref{alg:qnmfinder}. $t_{0}$ is the fitting start time initialized to be some $t_{0}^{f}$, with $t_{0}^{i}$ the earliest fitting start time; $N_{\mathrm{free}}$ is the number of free frequencies used in the \texttt{VarPro} fit initialized to be one; $\Delta t_{0}$ is some amount by which $t_{0}$ is changed after each failed iteration. More details about the algorithm are provided in Sec.~\ref{sec:codeoutline} and in the GitHub repository~\href{https://github.com/keefemitman/qnmfinder}{\textcolor{gray}{\faGithubSquare}}.
	}
	\label{fig:flowchart}
	\end{figure*}
}

\newcommand{\figGaussianTest}{%
	\begin{figure*}[tb]
		\includegraphics[width=\textwidth]{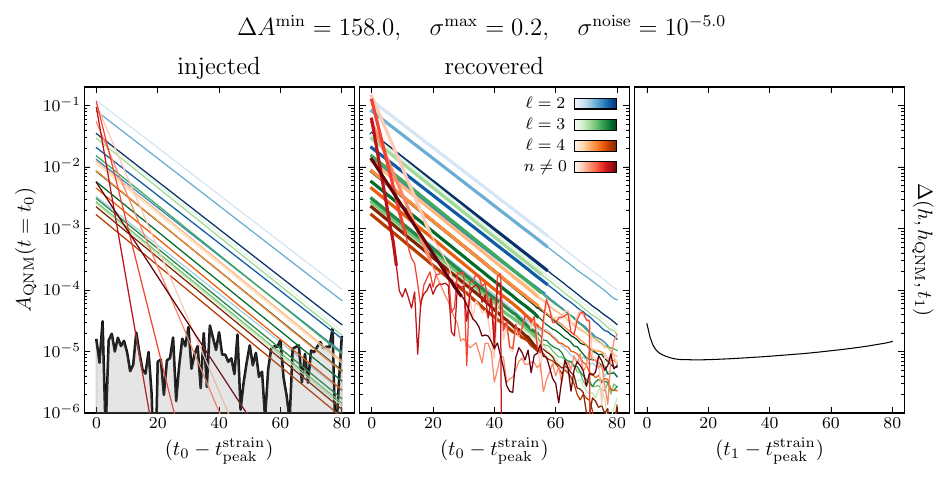}
		\caption{
			Analytic test with Gaussian noise on the order of $10^{-5}$, with the set of input QNMs matching those in Eq.~\eqref{eq:testQNMs}. Amplitudes are set by fitting a quasi-circular, $q=4$ precessing system at the time of peak strain $t_{\mathrm{peak}}^{\mathrm{strain}}$. \emph{Left}: QNM amplitudes as a function of time (colored) and the injected Gaussian noise (black). \emph{Middle}: QNM amplitudes as a function of time (colored); thin lines are the QNM model fit at every time, while the thick lines correspond to the time interval over which the QNM is most stable (see Eqs.~\eqref{eq:variationrequirement} and~\eqref{eq:timerequirement}). \emph{Right}: Residual error (see Eq.~\eqref{eq:residualerror}) between the model and the test waveform as a function of integration start time. Note that the error is on par with the magnitude of the injected Gaussian noise.
		}
		\label{fig:gaussian_test}
	\end{figure*}
}

\newcommand{\figNonGaussianTest}{%
	\begin{figure*}[tb]
		\includegraphics[width=\textwidth]{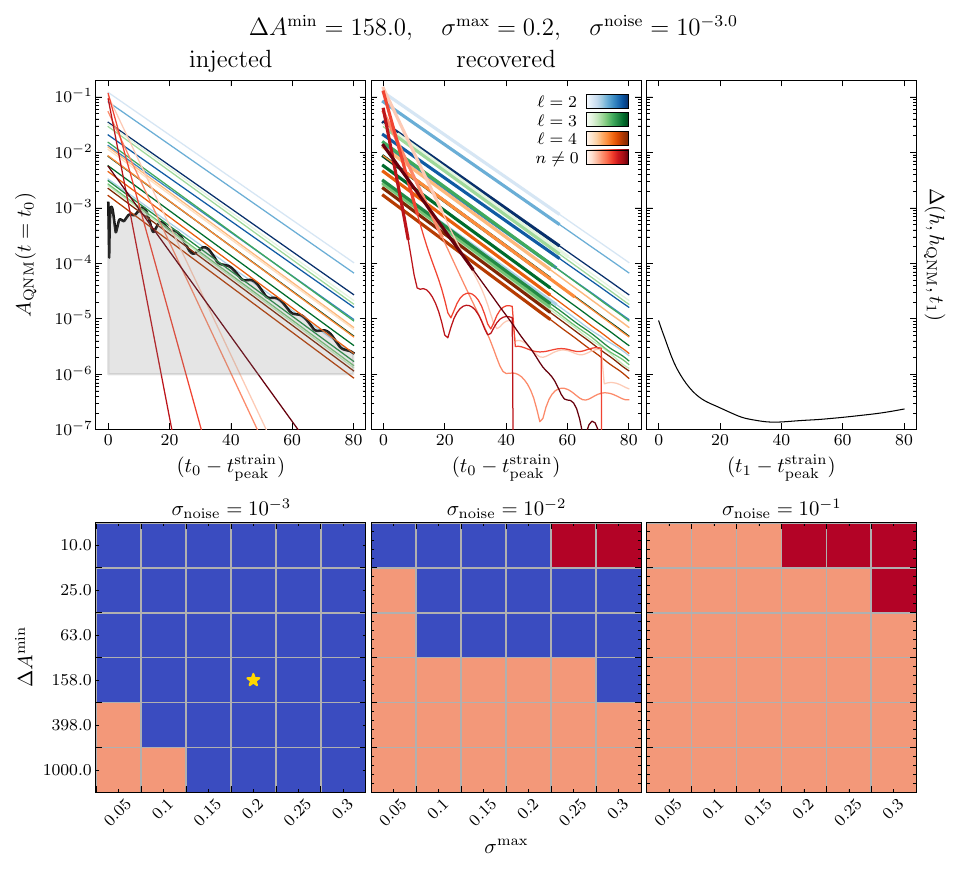}
		\caption{
			\emph{Top}: Identical to Fig.~\ref{fig:gaussian_test}, but instead with non-Gaussian noise inspired by the truncation error of NR simulations. \emph{Bottom}: Success rate of the algorithm versus different noise configurations as a function of different $\sigma^{\mathrm{max}}$ (see Eq.~\eqref{eq:variationrequirement}) and $\Delta A^{\mathrm{min}}$ (see Eq.~\eqref{eq:timerequirement}) tolerances. Blue squares correspond to every QNM being recovered, light red correspond to some of the QNMs being missed, and dark red correspond to spurious QNMs, i.e., QNMs not in the input waveform, being found as stable. A yellow star indicates the tolerances and noise that were used for the test shown in the top panel.
		}
		\label{fig:non_gaussian_test}
	\end{figure*}
}

\newcommand{\figNRExample}{%
	\begin{figure*}[tb]
		\includegraphics[width=\textwidth]{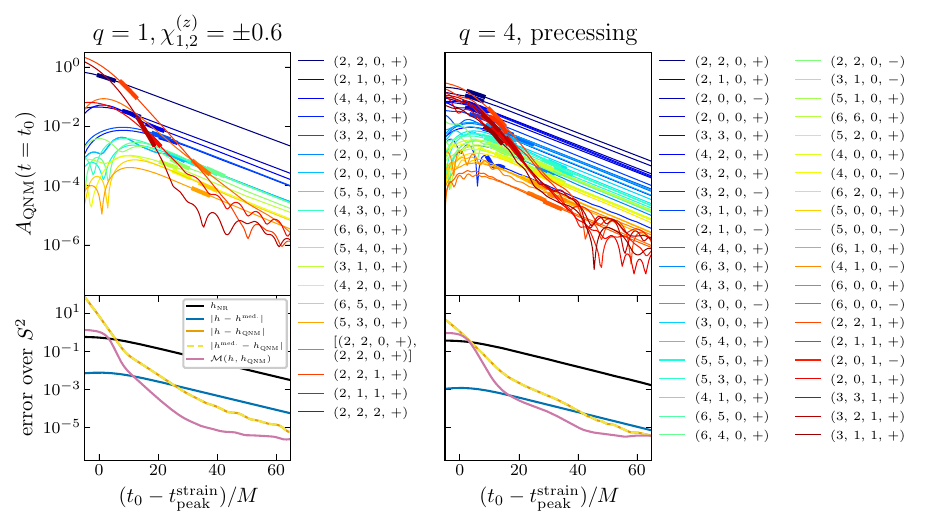}
		\caption{
			Applying the fitting algorithm to a $q=1$, $\chi_{1}^{(z)}=-\chi_{2}^{(z)}=0.6$ non-precessing system and a $q=4$, precessing system. Top panels show the QNM model fit across various start times; thin lines show the raw amplitudes, while thick lines show the amplitudes over the first $5M$ window in which they agree with the amplitude extracted over $\Delta t_{\mathrm{QNM}}$ within $10\%$---a rough proxy for the onset of the QNM's stability. The legend shows the QNMs in the order in which they are found. Bottom panels show various errors over the two-sphere; black is the NR waveform's norm, blue is the normed residual of the two highest resolutions, orange (yellow) is the normed residual of the (lower resolution) NR waveform and the QNM model, and magenta is the mismatch over the two-sphere (see Eq.~\eqref{eq:mismatch}) between the NR waveform and the QNM model.
		}
		\label{fig:nr_example}
	\end{figure*}
}

\newcommand{\figParameterSpaceQNMs}{%
	\begin{figure*}[tb]
		\includegraphics[width=\textwidth]{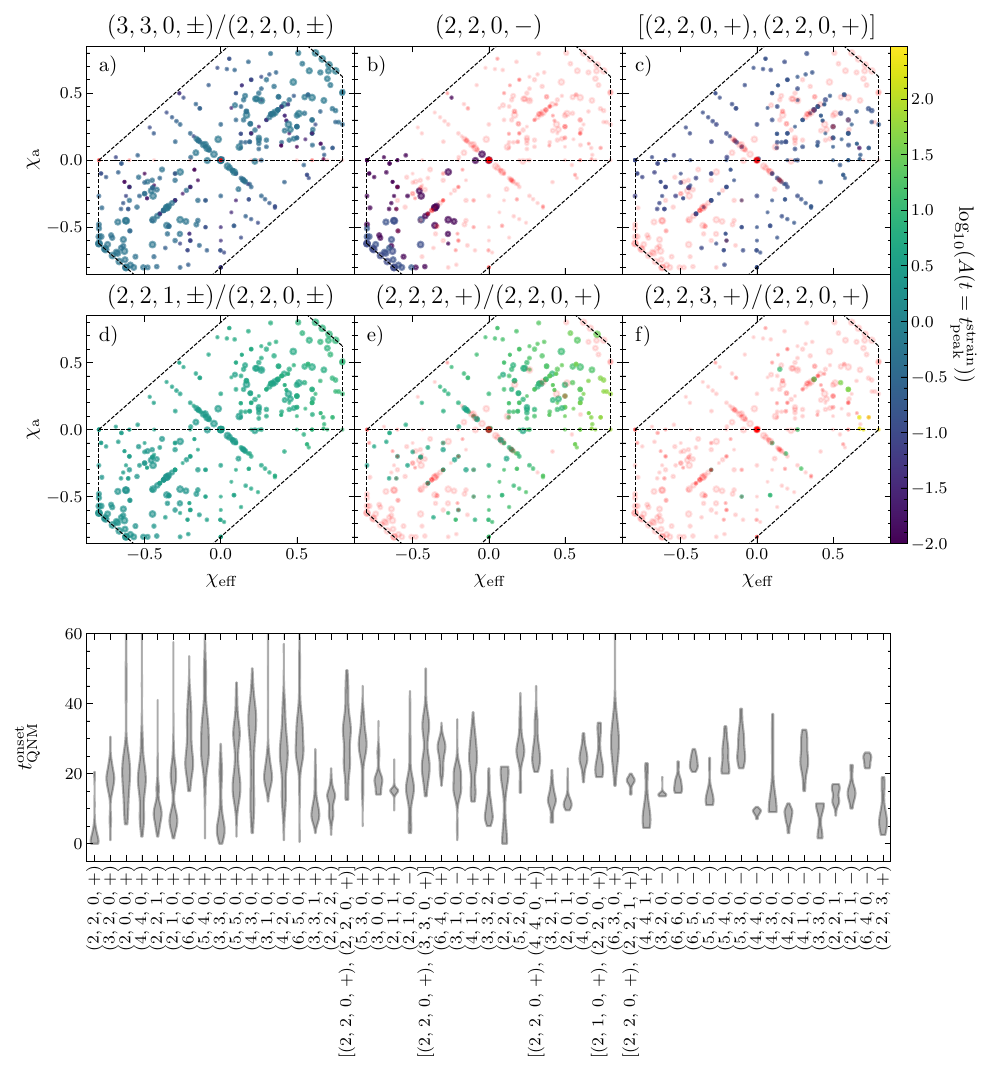}
		\caption{
			QNMs in quasi-circular, non-precessing parameter space that are recovered by the algorithm. The top panels show the QNM amplitudes (or amplitude ratios) as a function of $\chi_{\mathrm{eff}}$ and $\chi_{a}$ (see Eqs.~\eqref{eq:chieff} and~\eqref{eq:chia}) with a larger marker representing a larger mass ratio $q$. Red markers indicate simulations for which the QNM was not found by the algorithm. Note that in panels a) and d), if the remnant black hole is spinning in the negative $z$-direction, then we instead relabel retrograde QNMs as prograde for uniformity (see footnote 8 for our prograde/retrograde convention). The bottom panel shows the start time of the first $5M$ window in which the mean of the amplitude agrees with the amplitude extracted over $\Delta t_{\mathrm{QNM}}$ within $10\%$---a rough proxy for the onset of the QNM's stability---see Eq.~\eqref{eq:tonset}. The horizontal axis is ordered with respect to the number of times the QNM is found across the $\sim400$ simulations that were analyzed with the algorithm.
		}
		\label{fig:parameter_space_QNMs}
	\end{figure*}
}

\newcommand{\figParameterSpaceMismatches}{%
	\begin{figure*}[tb]
		\includegraphics[width=\textwidth]{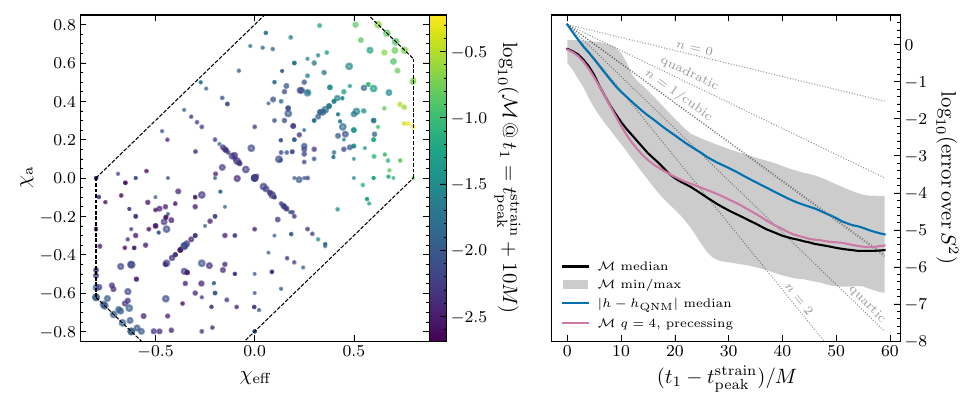}
		\caption{
			Mismatches between the QNM model built by the algorithm and the corresponding NR waveform across various points in quasi-circular, non-precessing parameter space. The left panel shows the mismatch at the time $t_{\mathrm{peak}}^{\mathrm{strain}}+10M$ (see Eq.~\eqref{eq:peakstrain}) as a function of $\chi_{\mathrm{eff}}$ and $\chi_{a}$ (see Eqs.~\eqref{eq:chieff} and~\eqref{eq:chia}) with marker size being the mass ratio $q$. The right panel shows various errors as a function of time. Black is the median mismatch, gray is the minimum and maximum values of the mismatch, blue is the median normed residual, and magenta is the mismatch of the $q=4$, precessing system that was shown in Fig.~\ref{fig:nr_example}. The dotted gray lines are included to show the decay rates of the $n=0,1$ and $2$ overtones as well as $n=0$ parent quadratic, cubic, and quartic QNMs, as a reference for the decay rate of the blue curve, which may indicate missing QNM content in our QNM model.
		}
		\label{fig:parameter_space_mismatches}
	\end{figure*}
}

\newcommand{\figOvertoneExcitation}{%
	\begin{figure*}[tb]
		\includegraphics[width=\textwidth]{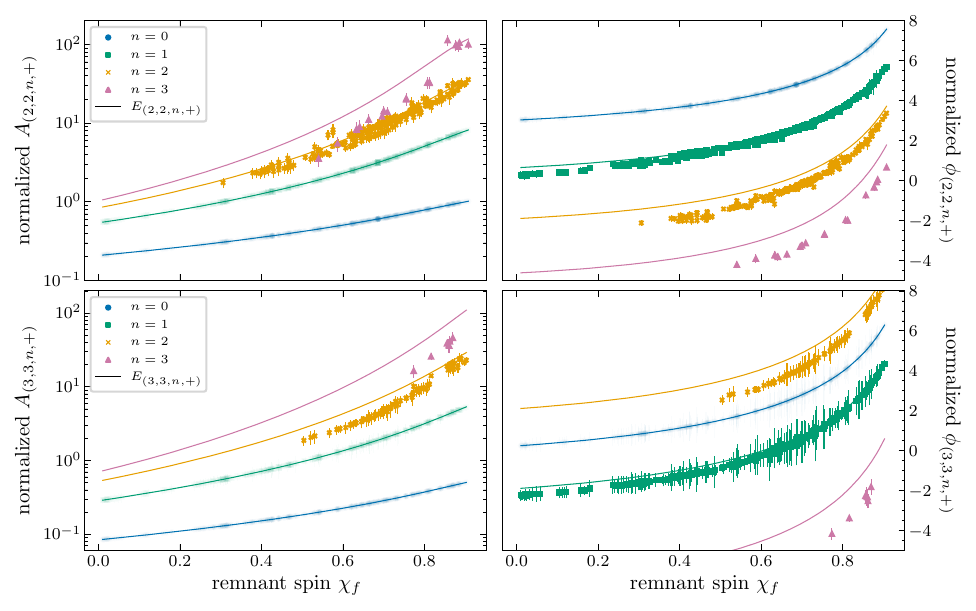}
		\caption{
			Normalized amplitudes and phases of the extracted $(2,2,n,+)$ QNMs, i.e., the right-hand side of Eq.~\eqref{eq:normalizedamplitudes}. Data points represent amplitudes extracted by the algorithm, while solid lines are the Teukolsky excitation factors computed in Refs.~\cite{Zhang:2013ksa,EFs1,EFs2} (scaled by a factor of $1/(-i\omega_{\mathrm{QNM}})^{2}$ so that they correspond to the strain rather than $\Psi_{4}$). Error bars are the standard deviation of the amplitude and phase over the QNM's most stable window. The $n=0$ and $n=1$ amplitudes and the $n=0$ phase are fixed to exactly match their excitation factors via Eqs.~\eqref{eq:EFparameters}, while every other data point is left unconstrained. This demonstrates that one can extrapolate to a time at which the amplitudes of the overtones nearly agree with their excitation factors, up to some constant. The phases follow the excitation factors to a similar degree, although larger deviations are observed for larger $n$. The bottom panel shows the same analysis as above, but for the $(3,3,n,+)$ QNMs with $(\ell,m)=(3,3)$ in Eqs.~\eqref{eq:EFparameters} instead of $(\ell,m)=(2,2)$, for which the trends are not as strong, though still somewhat smooth as a function of remnant spin.
		}
		\label{fig:overtone_excitation}
	\end{figure*}
}

\newcommand{\fitExcitationParameters}{%
	\begin{figure}[tb]
		\includegraphics[width=\columnwidth]{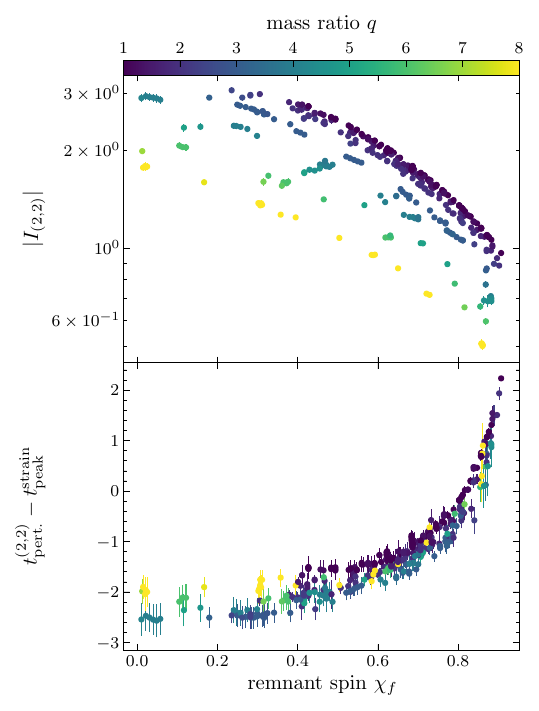}
		\caption{
			Values of Eqs.~\eqref{eq:EFparameters} needed to make the $n=0$ and $n=1$ amplitudes and the $n=0$ phase exactly match their excitation factors. Color represents the system's mass ratio $q$. Error bars denote the propagation of the QNM's amplitude and phase standard deviations through Eqs.~\eqref{eq:EFparameters}.
		}
		\label{fig:excitation_parameters}
	\end{figure}
}

\newcommand{\figQuadratic}{%
	\begin{figure*}[tb]
		\includegraphics[width=\textwidth]{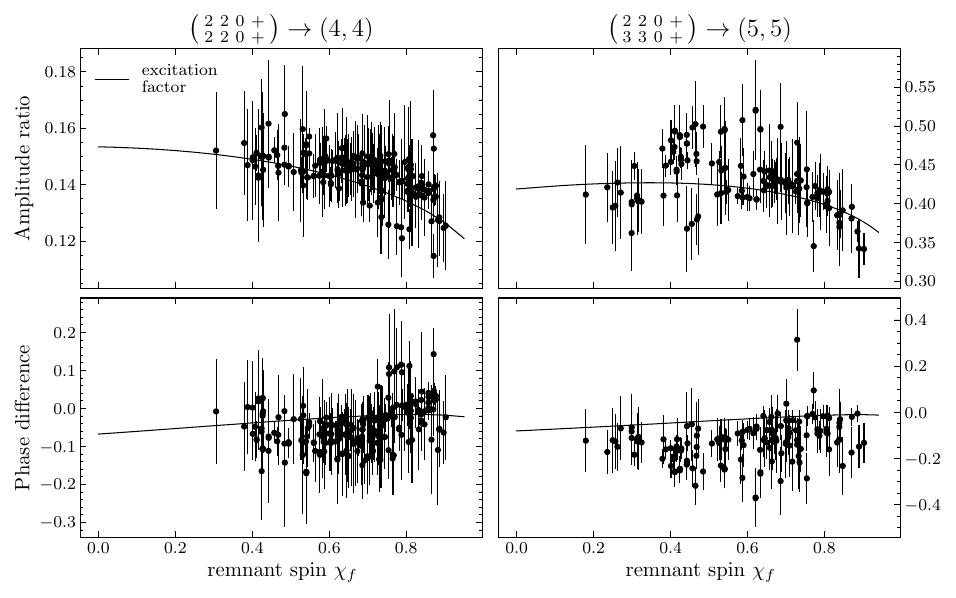}
		\caption{
			Quadratic $\begin{pmatrix}2&2&0&+\\2&2&0&+\end{pmatrix}\rightarrow(4,4)$ and $\begin{pmatrix}2&2&0&+\\3&3&0&+\end{pmatrix}\rightarrow(5,5)$ QNM amplitudes compared to their parent QNMs' amplitudes as a function of remnant spin. The top panel shows the relative amplitudes, while the bottom panel shows the absolute phase differences. Error bars represent one standard deviation over the QNM's stable window. In black we plot the excitation factor predicted by perturbation theory, i.e., that computed in Refs.~\cite{Ma:2024qcv,Khera:2024bjs}. Note that because this factor is computed for $M_{f}=1$, we scale the NR ratio by $1/M_{f}$, since each QNM's amplitude scales linearly with $M_{f}$.
		}
		\label{fig:quadratics}
	\end{figure*}
}

\newcommand{\figMemory}{%
	\begin{figure}[tb]
		\includegraphics[width=\columnwidth]{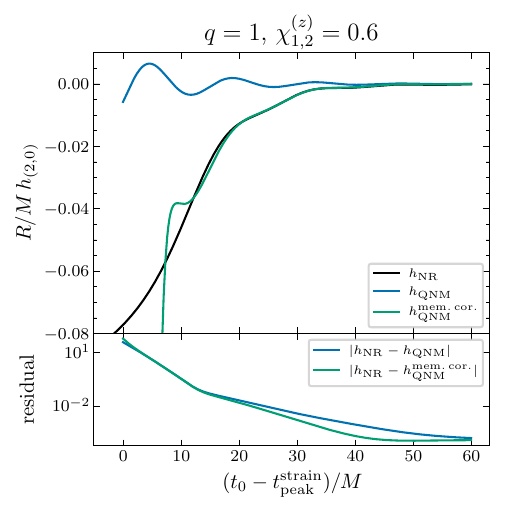}
		\caption{
			$(2,0)$ mode of the strain for a $q=1$, $\chi_{1,2}^{(z)}=0.6$ quasi-circular system. The top panel shows the NR waveform (black), the QNM waveform built by the algorithm (blue), and the memory-corrected QNM waveform (see Eq.~(17b) of Ref.~\cite{Mitman:2020bjf}) (green). The bottom panel shows the residuals, evaluated over the two-sphere via an $L^{2}$ norm.
		}
		\label{fig:memory}
	\end{figure}
}

\newcommand{\figGiesler}{%
	\begin{figure}[tb]
		\includegraphics[width=\columnwidth]{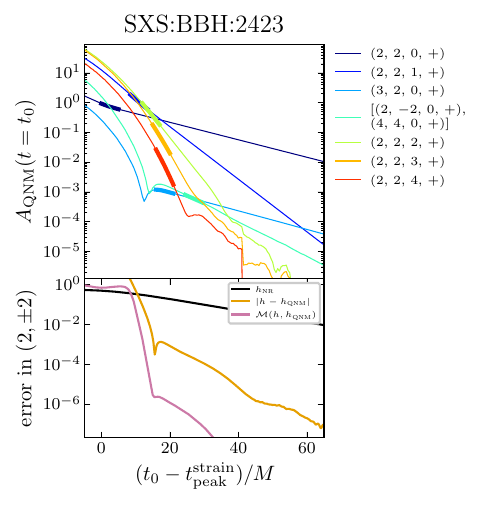}
		\caption{
			Identical to that shown in Fig.~\ref{fig:nr_example}, but for the SXS:BBH:2423 waveform studied in Ref.~\cite{Giesler:2024hcr}.
		}
		\label{fig:giesler}
	\end{figure}
}

\usepackage[T1]{fontenc}
\usepackage{lmodern}
\usepackage[utf8]{inputenc}
\usepackage{microtype}

\DeclareMathAlphabet{\mathbfsf}{\encodingdefault}{\sfdefault}{bx}{sl}

\usepackage[normalem]{ulem}

\newcommand{\CornellCcaps}{\affiliation{Cornell Center for Astrophysics and Planetary Science, Cornell University, Ithaca, New York 14853, USA}}

\newcommand{\Caltech}{\affiliation{Theoretical Astrophysics 350-17, California Institute of Technology, Pasadena, CA 91125, USA}}

\newcommand{\CCA}{\affiliation{Center for Computational Astrophysics, Flatiron Institute, New York NY 10010, USA}}

\newcommand{\Columbia}{\affiliation{
		Department of Physics, Columbia University,
		704 Pupin Hall, 538 West 120th Street, New York, New York 10027, USA
}}

\newcommand{\CornellPhysics}{\affiliation{Department of Physics, Cornell University, Ithaca, NY, 14853, USA}}

\newcommand{\Cornell}{\affiliation{Cornell Center for Astrophysics and Planetary Science, Cornell University, Ithaca, New York 14853, USA}}

\newcommand{\CornellLepp}{\affiliation{Laboratory for Elementary Particle Physics, Cornell University, Ithaca, New York 14853, USA}}

\begin{document}

	\hypersetup{pdfauthor={Mitman et al.}}
	
	\title{Probing the ringdown perturbation in binary black hole coalescences\\with an improved quasi-normal mode extraction algorithm}
	
	\author{Keefe Mitman\,\orcidlink{0000-0003-0276-3856}}
	\email{kem343@cornell.edu}
	\CornellCcaps
	\author{Isabella Pretto\,\orcidlink{0009-0001-7552-551X}}
	\Caltech 
	\author{Harrison Siegel\,\orcidlink{0000-0001-5161-4617}}
	\Columbia
	\CCA
	\author{Mark A. Scheel\,\orcidlink{0000-0001-6656-9134}}
	\Caltech
	\author{Saul A. Teukolsky\,\orcidlink{0000-0001-9765-4526}}
	\CornellCcaps
	\Caltech
	\author{\\Michael Boyle\,\orcidlink{0000-0002-5075-5116}}
	\Cornell
	\author{Nils Deppe\,\orcidlink{0000-0003-4557-4115}}
	\CornellLepp
	\CornellPhysics
	\Cornell
	\author{Lawrence E.~Kidder\,\orcidlink{0000-0001-5392-7342}} \Cornell
	\author{Jordan Moxon\,\orcidlink{0000-0001-9891-8677}}
	\Caltech
	\author{\\Kyle C.~Nelli\,\orcidlink{0000-0003-2426-8768}}
	\Caltech
	\author{William Throwe\,\orcidlink{0000-0001-5059-4378}} \Cornell
	\author{Nils L.~Vu\,\orcidlink{0000-0002-5767-3949}}
	\Caltech
	
	\begin{abstract}
		Using gravitational waves to probe the geometry of the ringing remnant black hole formed in a binary black hole coalescence is a well-established way to test Einstein's theory of general relativity. However, doing so requires knowledge of when the predictions of black hole perturbation theory, i.e., quasi-normal modes (QNMs), are a valid description of the emitted gravitational wave as well as what the amplitudes of these excitations are. In this work, we develop an algorithm to systematically extract QNMs from the ringdown of black hole merger simulations. Our algorithm improves upon previous ones in three ways: it fits over the two-sphere, enabling a complete model of the strain; it performs a reverse-search in time for QNMs using a more robust nonlinear least squares routine called \texttt{VarPro}; and it checks the variance of QNM amplitudes, which we refer to as ``stability'', over an interval matching the natural time scale of each QNM. Using this algorithm, we not only demonstrate the stability of a multitude of QNMs and their overtones across the parameter space of quasi-circular, non-precessing binary black holes, but we also identify new quadratic QNMs that may be detectable in the near future using ground-based interferometers. Furthermore, we provide evidence which suggests that the source of remnant black hole perturbations is roughly independent of the overtone index in a given angular harmonic across binary parameter space, at least for overtones with $n\lesssim2$. This finding may hint at the spatiotemporal structure of ringdown perturbations in black hole coalescences, as well as the regime of validity of perturbation theory in the ringdown of these events. Our algorithm is made publicly available at the following GitHub repository:~\href{https://github.com/keefemitman/qnmfinder}{\textcolor{gray}{\faGithubSquare}}.
	\end{abstract}
	
	\maketitle

	\section{Introduction}
	\label{sec:introduction}
	
	In the current era of gravitational wave (GW) science, Einstein's equations for vacuum general relativity (GR) have yet to be invalidated by any kind of consistency test. One of these tests which is particularly interesting is the no-hair test: an experiment to see whether an equilibrium black hole's mass and spin wholly characterize it~\cite{Echeverria:1989hg,Dreyer:2003bv,Berti:2005ys,Berti:2015itd,Berti:2016lat,Thrane:2017lqn,Baibhav:2017jhs,Baibhav:2018rfk,Isi:2019aib,CalderonBustillo:2020rmh}.\footnote{Black holes can also have electric charge, but this is expected to be negligible for astrophysical objects.} This experiment is conducted by studying a stage of binary black hole (BBH) coalescences called the ringdown ---the final stage of the coalescence in which the perturbed remnant black hole rings down to a state of equilibrium by emitting gravitational waves~\cite{Vishveshwara:1970cc,Press:1971wr,Teukolsky:1973ha,Chandrasekhar:1975zza}. This is because in the ringdown stage of a binary black hole coalescence there is a discrete spectrum of quasi-normal modes (QNMs) ---predicted by first-order black hole perturbation theory---that can be loudly excited by the perturbation sourced during the merger stage. So, by fitting damped sinusoids, i.e., QNMs, to observed data, one can check to see if the best fitting frequencies agree with the QNMs expected for some remnant mass and spin. If agreement is found, then the no-hair theorem is respected; if not, then vacuum GR is violated.\footnote{Note though that many objects which are not black holes may also generate QNMs, making interpretations of this test subtle~\cite{CardosoRingdownProbe_PhysRevLett.116.171101}.} However, there is another particularly useful feature of these QNM excitations besides their frequencies that can be used to test GR: their \emph{amplitudes}.
	
	Just like how a certain mass and spin constrain what type of QNM frequencies one should expect in ringdown, the progenitor parameters in a binary black hole merger also constrain the relative excitation\footnote{Examining the relative excitation is key because the magnitudes of the complex QNM amplitudes are degenerate with the binary's total mass and luminosity distance.} of QNMs~\cite{Berti:2007fi,Kamaretsos:2011um,Kamaretsos:2012bs,London:2014cma,London:2018gaq,Borhanian:2019kxt,JimenezForteza:2020cve,Cheung:2023vki,Pacilio:2024tdl,MaganaZertuche:2024ajz,Gao:2025zvl}. Consequently, one can also perform a test of GR by seeing whether or not the relative QNM amplitudes for some BBH match what GR, i.e., numerical relativity (NR), predicts for the BBH progenitor parameters' posterior. Or, by performing the inverse analysis, one could even infer BBH progenitor parameters from the QNMs alone, which may provide valuable astrophysical information related to binary formation channels. However, to do so requires that one can confidently extract QNM amplitudes from numerical simulations, and this task has proven to be challenging thus far~\cite{Baibhav:2023clw,Zhu:2023mzv,Nee:2023osy,Cheung:2023vki,Nee:2023osy,Takahashi:2023tkb,Clarke:2024lwi,Giesler:2024hcr}.
	
	This extraction is troublesome for the following reason: what does it mean for a QNM that has been extracted from an NR waveform to be physically meaningful and not simply the result of over fitting to some feature in the data that is not a damped sinuoid with the QNM's frequency? This concern arises because while QNMs are expected to contain much of the power in the ringdown stage of BBHs, first-order black hole perturbation theory also predicts the presence of a prompt response---the propagation of the perturbation directly to the observer~\cite{Leaver:1986gd,Nollert:1999ji}---and a series of power law tails---the manifestation of back-scattering off of the black hole's potential~\cite{DeWitt:1960fc,Ma:2024hzq,DeAmicis:2024eoy}---that are not simple damped sinusoids, like the QNMs. Furthermore, apart from the non-QNM content that perturbation theory predicts, between the merger and ringdown stages it is also possible that there will be some non-perturbative content present in the waveform. Thus, extracting QNMs from NR waveforms---which may also have some degree of numerical noise---requires extreme caution to ensure that the extraction is performed robustly.
	
	One way to ensure that the QNMs extracted from an NR waveform are physical is by requiring that they satisfy some stability criterion with respect to the time of the QNM extraction~\cite{London:2014cma,Baibhav:2023clw,Cheung:2023vki}. That is, over some interval of subsequent extraction times, the QNM amplitudes that one obtains at some time should track the evolution of the damped sinusoid (or, equivalently, the QNM amplitudes projected to some reference time should be constant).\footnote{Note, however, that even first-order perturbation theory predicts that the QNMs are not always stable as a function of time~\cite{Andersson:1996cm,Chavda:2024awq}. This is because the perturbation that sources the QNMs is not a delta function, but instead has some spatiotemporal profile that yields a time-dependent excitation. And, since this profile of BBH remnant perturbations remains to be well understood, we will not try to account for it in the remainder of this work.} On the other hand, if non-QNM content is mistakenly modeled as a QNM, then the fitted amplitude should not track the expected evolution. So, if stability can be shown, then one can be more confident that the QNM amplitude that they extracted really represents some QNM in the data, rather than some non-QNM feature that the QNM happens to have some non-negligible correlation with.
	
	The main reason why showing QNM stability has been particularly onerous was clearly demonstrated in Ref.~\cite{Giesler:2024hcr}: fitting damped sinusoids is unfortunately an inherently ill-conditioned procedure~\cite{Nee:2023osy,Baibhav:2023clw,Zhu:2023mzv,Giesler:2024hcr}. A small amount of noise in the data produces a much larger effect on the values of the fitted parameters. Nevertheless, in Ref.~\cite{Giesler:2024hcr} the vital task of showing QNM stability, especially for the faster-decaying QNMs, was achieved. However, even though the stability of many QNMs was shown, finding the QNMs that could be confidently extracted from the NR waveforms was a rather hands-on exercise. And, if QNM excitations are to be studied across all of BBH parameter space (which is important for performing this test of GR based on the QNM amplitudes), then an automated algorithm for extracting whatever QNMs exist in the corresponding NR waveforms is needed.
	
	So far, there have been a few attempts to automate QNM extraction from NR waveforms by using various types of greedy algorithms~\cite{London:2014cma,London:2018gaq,MaganaZertuche:2021syq,Cheung:2023vki}, with the most similar to this work being the one presented in Ref.~\cite{Cheung:2023vki}. In Ref.~\cite{Cheung:2023vki} the algorithm that they presented effectively consisted of the following. For a spherical harmonic mode of the NR waveform, perform an agnostic frequency search with some number of free frequencies to find QNMs that may be in the data. Once certain QNMs' frequencies are identified, fit the corresponding damped sinusoids to the NR waveform at varying start times and check to see if their amplitudes are stable over some interval. If not, remove those QNMs from the QNM model until only a set of stable QNMs remains. Then, take the best model to simply be the model with the largest number of stable QNMs. Using this method, Ref.~\cite{Cheung:2023vki} was able to extract many of the longest lived QNMs, some of the 2nd longest lived QNMs, and even certain ``quadratic'' QNMs expected by 2nd-order perturbation theory~\cite{London:2014cma,Mitman:2022qdl,Cheung:2022rbm}. However, their algorithm, as well as the other algorithms that have been presented in the literature, was unable to identify many of the faster-decaying QNMs that can contain a large fraction of the waveform's power at earlier times in the ringdown~\cite{Giesler:2024hcr}. Therefore, to help understand the QNM content across BBH parameter space and thus prepare us for performing ringdown-based tests of GR on observed gravitational wave data, there is a pressing need for an automated routine that can reliably, and meaningfully, extract any faster-decaying QNM content from the ringdown stage of NR simulations.
	
	In this work we present such an algorithm. In particular, we improve upon previous works in three ways. First, we focus on building a QNM model that best models the ringdown stage of a BBH over the entire two-sphere, instead of only in one $(\ell, m)$ harmonic mode~\cite{Cook:2020otn,MaganaZertuche:2021syq,Gao:2025zvl}. This is important because for different types of BBHs, $\ell=2$ modes besides the $(2,2)$ mode and higher $\ell$ modes can contain a large fraction of the emitted power~\cite{Zhu:2023fnf}. Second, we also utilize the fact that the QNMs have different decay times to identify the longest-lived QNMs first using the late portion of the ringdown, and then the faster-decaying QNMs by studying the NR waveform at progressively earlier times. And, finally, we introduce a QNM-dependent stability criterion that is much more representative of the stability we expect to see in theory,\footnote{For details on what we mean by this, see Sec.~\ref{sec:codeoutline}.} and therefore are able to resolve faster-decaying QNMs that previous algorithms could not. These last two points in particular set us up to identify QNMs which were previously inaccessible with previous extraction methods. However, we stress that fitting over the two-sphere also enables us to markedly narrow down the space of expected quadratic QNMs, which proves to be key for excluding spurious QNMs (for more on this, see Sec.~\ref{sec:codeoutline}).\footnote{We would also like to point out that Ref.~\cite{Giesler:2024hcr} utilized many of the techniques we use here, such as searching for frequencies with \texttt{VarPro} at progressively earlier times. However, we stress that our work is unique since no algorithm to automate this procedure was developed in Ref.~\cite{Giesler:2024hcr}.}
	
	We demonstrate the algorithm's accuracy by testing it against a series of noisy mock examples for which the QNMs are known. Unlike previous works, we not only study tests with Gaussian noise, but also tests with non-Gaussian noise inspired by numerical relativity noise. Once the algorithm is validated against these various analytic tests, with a large assortment of high-accuracy NR simulations we highlight the algorithm's ability to not only extract previously identified QNMs, but also a vast range of the faster-decaying QNMs, as well as some quadratic QNMs that have not been identified before. Finally, by applying the algorithm to $400$ quasi-circular, non-precessing simulations, we significantly improve upon previous relative QNM excitation measurements and even introduce new ones that yield interesting information about the nature of ringdown perturbations in BBHs.
	
	The outline of the paper is as follows. In Sec.~\ref{sec:algorithm} we present the algorithm that we develop and use to extract QNMs from NR waveforms. In Sec.~\ref{sec:codeoutline} we present a detailed overview of the code, while in Sec.~\ref{sec:analytictests} we validate the algorithm against a series of analytic tests for which the QNM content is known. In this section we also provide motivation for the tolerances required by the algorithm that we use when studying real NR waveforms. In Sec.~\ref{sec:results} we present the QNMs found by the algorithm for a series of NR simulations. In Sec.~\ref{sec:NRwaveforms} we introduce the various NR simulations whose ringdowns we analyze with the algorithm as well as how we pre-process the waveforms to be in the canonical frame expected by perturbation theory. In Sec.~\ref{sec:NRfits} we present what the algorithm manages to extract from example NR simulations. Then, in Sec.~\ref{sec:QCNPsystems} we illustrate what the algorithm manages to find across the quasi-circular, non-precessing part of parameter space. In particular, we show that the algorithm manages to obtain post-peak mismatches well below the requirements of the current LVK detectors in addition to resolving many of the faster-decaying QNMs. Lastly, in Sec.~\ref{sec:EFanalysis}, we present new results relating QNM amplitudes to their excitation factors and new analyses of quadratic QNMs that may be detectable in the near future. Our results provide key insights into the excitation of QNMs and the nature of ringdown perturbations in BBHs. In Sec.~\ref{sec:discussion} we then conclude by summarizing our algorithm and findings and discussing potential future applications. In App.~\ref{app:reproduce}, we also compare the QNMs that our algorithm extracts from the simulations analyzed in Ref.~\cite{Giesler:2024hcr} to the QNMs found in that work and comment on their differences.
	
	Readers who are only interested in the physics that we manage to extract with the algorithm, but not the details of the algorithm itself, should skip to Sec.~\ref{sec:NRfits}.
	
	\section{Extraction Algorithm}
	\label{sec:algorithm}
	
	As outlined in Sec.~\ref{sec:introduction}, the final state of a BBH merger is a perturbed Kerr solution that settles by emitting radiation in the form of gravitational waves (GWs)~\cite{Vishveshwara:1970cc,Press:1971wr,Teukolsky:1973ha,Chandrasekhar:1975zza}. According to black hole perturbation theory, these GWs have three unique contributions: the prompt response (which dominates at very early times), QNMs (which dominate at intermediate times), and power-law tails (which dominate at late times)~\cite{Teukolsky:1973ha,Leaver:1986gd,Nollert:1999ji,DeWitt:1960fc,Ma:2024hzq,DeAmicis:2024eoy}. Unfortunately, very little is known about the morphology of the prompt response in binary black hole mergers,\footnote{See Ref.~\cite{Andersson:1995zk} for details on the morphology of the prompt response in scattering simulations and Ref.~\cite{Chavda:2024awq} for some insightful analyses of more complicated test problems.} whereas the QNM and power-law tail contributions to the ringdown have been studied extensively. In this work we exclusively focus on the QNM contribution, since for astrophysical BBHs this is expected to exhibit the most power and thus be the most detectable part of the ringdown stage.
	
	After the prompt response, and once the excitation of the QNMs has finished~\cite{Andersson:1995zk,Andersson:1996cm,Chavda:2024awq}, the emitted GW strain $h$ as a function of time $t$ and angular coordinates on the two-sphere $\theta,\phi$ can be expressed as an infinite sum of damped sinusoids
	\begin{align}
		\label{eq:QNMwaveform}
		h(t,\theta,\phi)=\sum\limits_{k=\left\{\substack{\ell\geq2,|m|\leq\ell,\\n\geq0,p=\pm1}\right\}}C_{k}e^{-i\omega_{k}t}\phantom{}_{-2}S_{(\ell,m)}(a\omega_{k})(\theta,\phi),
	\end{align}
	where $(\ell,m)$ are angular harmonic numbers, $n$ is the overtone number (with $n=0$ being the longest-lived, fundamental mode), $p$ is the prograde/retrograde index,\footnote{In this work we take  $p=\mathrm{sgn}(m\mathrm{Re}[\omega])$, so $p=+$ corresponds to a prograde QNM while $p=-$ is a retrograde QNM. For the special $m=0$ QNMs, there is no longer a connection to prograde or retrograde phase fronts so we instead take $p=\mathrm{sgn}(\mathrm{Re}[\omega])$.} $C_{(\ell,m,n,p)}$ is some complex amplitude, $\omega_{(\ell,m,n,p)}$ is the QNM frequency, and $\phantom{}_{-2}S_{(\ell,m)}(a\omega)$ is the spin-weight $-2$ spheroidal harmonic with $a$ the dimensionful spin of the black hole (see, e.g., Ref.~\cite{Berti:2009kk} for an in-depth review). Again, we stress that in Eq.~\eqref{eq:QNMwaveform}, the frequencies $\omega$---according to the no-hair theorem---are completely fixed by the remnant black hole's mass $M_{f}$ and spin $a$, while the complex amplitudes $C$ are unknown functions of the BBH's progenitor parameters and can only be determined by extracting them from NR simulations. This extraction is often performed using linear least squares because, for a fixed set of frequencies, the complex amplitudes are nothing more than linear parameters in the QNM fit. Furthermore, because NR waveforms are often expressed in a spin-weight $-2$ spherical harmonic basis (rather than the spheroidal harmonic basis used in Eq.~\eqref{eq:QNMwaveform}), when fitting them one often projects the spheroidal harmonic in Eq.~\eqref{eq:QNMwaveform} onto spherical harmonics using the well-known spherical-spheroidal mixing coefficients~\cite{Stein:2019mop}. As a result, within a specific $(\ell, m)$ spherical harmonic mode of an NR waveform, one may also find the $(\ell'\not=\ell,m)$ QNMs, e.g., the $(3,2,0,+)$ QNM can be in the $(2,2)$ mode. 
	
	\figFlowChart
	
	\algQNMFinder
	
	However, even though fitting for these amplitudes is fairly straight-forward it is not known, a priori, exactly which QNMs may be sourced in a BBH merger, so it is also important to have a procedure that can fit for some number of free frequencies. Nonlinear least squares routines for fitting frequencies are less prone to issues than other commonly used methods. The ``best'' algorithm for doing this is known and is called variable projection, or \texttt{VarPro} for short~\cite{golub1972,golub1973}. \texttt{VarPro} works by performing an iterative nonlinear fitting routine; first, solve for the nonlinear parameters, i.e., the frequencies, using some initial guess, then use standard linear least squares to solve for the linear parameters with fixed nonlinear parameters, and finally fix these linear parameters and perform the nonlinear fit once more, repeating this process until a suitable tolerance is achieved.\footnote{We find that \texttt{VarPro} is very insensitive to the initial guess. So, rather than run \texttt{VarPro} at many randomly chosen initial guesses which will decrease the algorithm's efficiency, we simply take the initial guess to be $a-ib$, where $a=0.25((n\%3)-2)(3(n\%3)-1)$ and $b=0.1\lceil (n+1)/3\rceil$, where $n$ is the $n^{\mathrm{th}}$ free frequency, i.e., the $n^{\mathrm{th}}$ initial guess. This corresponds to taking the real part to be $0.5$, $-0.5$, or $0$, and the imaginary part to be a multiple of $-0.1$. Note that the \% operator represents the modulo operation.} During this procedure, note that the Jacobian of the model with respect to the nonlinear parameters is needed. But, because the sum of damped sinusoids is a simple analytic function, obtaining this is not particularly challenging. Consequently, \texttt{VarPro} is an important tool for identifying whatever QNMs may be in the data. In this work, we use \texttt{VarPro} extensively.
	
	As stressed earlier, though, fitting damped sinusoids with fixed or free frequencies is not enough to solve the problem of identifying the QNM content in the ringdown of a BBH system. Instead, one must also show that the extracted content matches what one expects physically, i.e., that it is stable across extraction times, provided one is sufficiently late in the ringdown. We outline our procedure to do this in Sec.~\ref{sec:codeoutline}. While we summarize the main idea here in the text, we also provide a flow chart in Fig.~\ref{fig:flowchart} as well as an algorithmic description of the routine in Alg.~\ref{alg:qnmfinder} for those that may find these more insightful.
	
	\subsection{Code outline}
	\label{sec:codeoutline}
	
	Our extraction method is based on the following idea. At late times, only the longest-lived QNMs should be present in relation to whatever the truncation error of the NR simulation is. Therefore, one should construct the QNM model iteratively, identifying as many QNMs as possible at late times before moving to earlier times to try to identify faster-decaying QNMs, like overtones and quadratic QNMs. But, even at late times, a priori the QNM content is not exactly known. So, to begin with, one should use \texttt{VarPro} to fit some free frequency and try to match it to some QNM corresponding to the remnant black hole's mass and spin. In principle, one could search for the best-fitting frequency over the entire two-sphere. But in practice, because the current implementation of \texttt{VarPro} can only search for frequencies in one harmonic at a time, we instead utilize the following. For iteration with time $t_{0}$ (moving from late times to early times), compute the power in the residual between the NR waveform and the QNM model for each spherical harmonic mode, i.e.,
	\begin{align}
		\label{eq:unmodeledpower}
		\mathcal{E}_{(\ell, m)}=\int_{t_{0}}^{t_{f}}\int_{S^{2}}Y_{(\ell,m)}^{*}|\dot{h}_{\mathrm{NR}}-\dot{h}_{\mathrm{QNM}}|^{2}d\Omega\,dt,
	\end{align}
	where $\phantom{}^{*}$ represents complex conjugation and an over-dot represents a time derivative~\cite{Ruiz:2007yx}. Of these spherical modes, find the one with the most power---this will be the worst modeled mode in the NR waveform. In this mode, use \texttt{VarPro} to find some best-fitting frequency with respect to whatever the QNM model is. Then, iterate over the possible QNMs that could exist within this mode and find the QNM whose frequency is the closest to the frequency returned by \texttt{VarPro}. Once this potential QNM is found, one can then proceed to check if it (and the whole model) is stable with respect to different fitting start times.
	
	In previous works, QNM stability criteria have largely consisted of requiring the fractional variation in some combination of the amplitude and phase to be below some tolerance over some time interval of a predetermined size, e.g., $10M$, where $M$ is the total Christodoulou mass of the two black holes~\cite{Christodoulou:1971pcn}. In this work, we will use a similar error measure, but will instead consider the variation of point estimates of the complex amplitudes' real and imaginary parts at different fitting times:\footnote{Eq.~\eqref{eq:variation} is not an explicit error of the measured parameters, since it is only tracking changes in point estimates over the fit extraction times, rather than statistical uncertainty in the fit at each time. However, since our fits are dominated by systematics, the fit's statistical uncertainties are subdominant and can be safely ignored for the purposes of this work.} 
	\begin{align}
		\label{eq:variation}
		\sigma_{\mathrm{QNM}}\equiv\sqrt{\left(\frac{\Delta\mathrm{Re}\left[C_{\mathrm{QNM}}\right]}{|C_{\mathrm{QNM}}|}\right)^{2}+\left(\frac{\Delta\mathrm{Im}\left[C_{\mathrm{QNM}}\right]}{|C_{\mathrm{QNM}}|}\right)^{2}},
	\end{align}
	where $C_{\mathrm{QNM}}$ is the QNM's complex amplitude measured at a reference time, which we take to be the time of the strain's peak over the two-sphere
	\begin{align}
		\label{eq:peakstrain}
		t_{\mathrm{peak}}^{\mathrm{strain}}\equiv\mathrm{argmax}_{t}\left(\int_{S^{2}}|h(t,\theta,\phi)|^{2}d\Omega\right).
	\end{align}
	The $\Delta$ in Eq.~\eqref{eq:variation} represents the standard deviation of the  following quantity over the relevant time interval. However, unlike previous studies, we will not require a QNM to have
	\begin{align}
		\label{eq:variationrequirement}
		\sigma_{\mathrm{QNM}}<\sigma^{\mathrm{max}}
	\end{align}
	for some $\sigma^{\mathrm{max}}$ tolerance over a window of some \emph{fixed} size. Instead, we will require that a QNM obey Eq.~\eqref{eq:variationrequirement} only over the time window in which its amplitude decays by some fixed amount, provided that fixed amount is above some arbitrary minimum which we take to be $4M$,\footnote{In practice, the algorithm never searchers for a QNM over a window as short as $4M$, which is part of why we choose this value. But this will likely need to be revisited in the future if even faster-decaying QNMs can be identified.} i.e.,
	\begin{align}
		\label{eq:timerequirement}
		\Delta t_{\mathrm{QNM}}=\mathrm{max}(-\frac{\ln\left(\Delta A^{\mathrm{min}}\right)}{\mathrm{Im}\left[\omega_{\mathrm{QNM}}\right]}, 4)
	\end{align}
	for some $\Delta A^{\mathrm{min}}$ tolerance.\footnote{Note that in Eq.~\eqref{eq:timerequirement} $\Delta A^{\min}$ represents the factor by which the amplitude changes, e.g., $\Delta A^{\min}=100$ yields a time window over which the QNM decays by a factor of $100$. For information on how we choose values for $\sigma^{\mathrm{max}}$ and $\Delta A^{\mathrm{min}}$, see Sec.~\ref{sec:analytictests}.} We require stability this way because it is more motivated by the problem at hand. Specifically, every QNM has its own time scale set by its damping time---$\mathrm{Im}\left[\omega_{\mathrm{QNM}}\right]$---so it should only be required to be stable over a time window related to this time scale, e.g., Eq.~\eqref{eq:timerequirement}.\footnote{It is easy to verify that the Fisher matrix for a single QNM only depends on $\mathrm{Im}[\omega_{\mathrm{QNM}}]$. While the real part is indeed relevant for Fisher matrices involving more than one QNM, its impact on the matrix is still subdominant to the imaginary part, which is why we only utilize the imaginary part in Eq.~\eqref{eq:timerequirement}.} This consideration is important because if the interval is too large and the QNM decays to the noise floor, then it may be deemed unphysical. And if the interval is too small, then statistical significance is lost\footnote{Since the number of point estimates that can be used decreases.} and spurious QNMs may be found. We reiterate that this version of the stability criterion is a crucial adaptation to ensure that faster-decaying QNMs, like overtones and quadratics, can be robustly extracted from NR waveforms. Consequently, once we find a QNM that matches (within some tolerance) \texttt{VarPro}'s best-fitting frequency, we check the stability of that QNM and the other QNMs in the model by requiring that they obey Eq.~\eqref{eq:variationrequirement} over their own independent time interval computed via Eq.~\eqref{eq:timerequirement}. 
	
	If a QNM passes our stability criterion, then we add it to the model and continue searching at late times. However, if a QNM does not pass, then we move to an earlier time and run \texttt{VarPro} again to try to find some other prospective QNM. If at some point we reach a time where the mode with the most unmodeled power via Eq.~\eqref{eq:unmodeledpower} changes, then rather than searching in the new mode, we instead repeat the search in the previous mode, but with one additional free frequency in the \texttt{VarPro} fit. This is essential because it is often the case that there are many QNMs with similar amplitudes, so trying to verify the stability of one at a time is practically impossible. One obvious example of this is trying to extract QNMs from $m=0$ modes in non-precessing systems, for which the two $p=\pm$ mirror mode amplitudes are identical, up to complex conjugation. But we stress that there can be even more complicated combinations of QNMs for which this routine is vital for stably extracting them. We continue this procedure until we reach some preset maximum number of free frequencies $N^{\mathrm{max}}$, at which point we continue with the search at earlier times in some other poorly modeled spherical harmonic mode. We find that $N^{\mathrm{max}}=4$ is sufficient for the cases studied.
	
	Finally, one other important feature that we introduce into the algorithm is that, whenever a QNM is searched for to match whatever frequencies are returned by \texttt{VarPro}, we require that the overtones of some $(\ell, m, \cdot, p)$ QNM are added sequentially. That is, for overtone $n=N$ to be considered a potential QNM, overtone $n=N-1$ must already be in the model or be under consideration if multiple frequencies are being examined. We also employ a similar requirement for quadratic QNMs. In particular, we require that their parent modes must be in the model and their amplitude must be less than the product of their parent modes' amplitudes\footnote{For this we really require that the quadratic QNM's amplitude is less than $10$ times the product of their parent modes' amplitudes, just to make this requirement less restrictive than it could be.} to be considered as a potential model addition. The first of these two conditions is well-motivated since each successive overtone has a faster decay time and thus will be identified by \texttt{VarPro} after any lower-$n$ overtones. The second condition is also motivated since the quadratic QNMs' excitation factors are typically $\lesssim\mathcal{O}(0.1)$~\cite{Ma:2024qcv} and thus would require one parent QNM to have an extremely large amplitude relative to the other for the quadratic QNM to have more power than one of the parent QNMs. Ideally, though, we would not even search for quadratic QNMs, but would include them automatically once their parent modes are found using the quadratic QNM ratios, such as those computed in Ref.~\cite{Ma:2024hzq}. But, since these ratios do not exist for all possible combinations, we reserve this for future work. Overall we find these two conditions to be helpful in limiting the number of QNMs that need to be searched for to match some frequency returned by \texttt{VarPro}.
	
	Apart from this, there a few other subtleties involved with the algorithm that we list now. But the uninterested reader may skip to Sec.~\ref{sec:analytictests} to see the validation of the algorithm against analytic tests. In addition to everything described above, the algorithm also has the following features to aid with accuracy and efficiency:
	\begin{enumerate}
		\item When searching for poorly-modeled modes via Eq.~\eqref{eq:unmodeledpower}, if $\mathcal{E}_{(\ell,m)}$ is less than some amount, i.e., the mode has very little unmodeled power, then we simply skip that mode and continue the search at earlier times until we find a mode with a more significant amount of unmodeled power. We take this power tolerance value to be $10^{-12}$, as we find this to be consistent with the power one obtains from the constant late-time numerical noise floor.
		\item When searching for the QNMs that best match the best-fitting frequency returned by \texttt{Varpro}, if the magnitude of the complex frequencies' difference is above some tolerance then we do not consider a match to be found and continue the search with \texttt{VarPro} at some earlier time. We find that our results are quite independent of this tolerance and we set it to be a modest value of $\Delta\omega^{\mathrm{max}}=0.2$.
		\item Whenever an $(\ell,m,n,p)$ QNM is found to be stable and is added to the model, we immediately check to see if its mirror, $(\ell,-m,n,p)$, can also be added to the model as a stable QNM. This is because these two QNMs are excited to equivalent degrees in non-precessing systems, and very similar degrees in precessing systems. Note that the amplitude of the mirror mode is left independent of the original to account for these asymmetries in precessing systems.
		\item For the analytic tests in Sec.~\ref{sec:analytictests}, the faster-decaying QNMs can decay to be exactly zero at late times and thus be found to be stable (even though their values are meaningless). Consequently, we require that each QNM be stable not only over $\Delta t_{\mathrm{QNM}}$, but also in the regime where $A>A^{\mathrm{min}}=10^{-12}$.
		\item Because the angular mixing of the quadratic QNMs into higher $\ell$ modes is not currently available~\cite{Ma:2024qcv}, we do not fit quadratic QNMs over the two-sphere, but instead only in the mode in which they were found by \texttt{VarPro}. This means that we could in principle fit the $(2,2,0,+)\times(2,2,0,+)$ quadratic QNM independently to the $(4,4)$ and $(5,4)$ modes, however, this never happens in practice since the algorithm only ever finds the quadratic QNMs to be stable in the most-dominant harmonic. This can be replaced in the future once all of the quadratic QNM ratios have been computed.
		\item For the $m=0$ modes, quadratic QNMs related to memory effects (see, e.g., Ref.~\cite{Mitman:2024uss} for a review) can appear with $\mathrm{Re}\left[\omega_{\mathrm{QNM}}\right]=0$, i.e., those stemming from $(\ell,m,n,p)\times(\ell,-m,n,p)$. As a result, it is incredibly challenging to distinguish these quadratic QNMs from one another since the imaginary part of their frequencies can be very similar. Fortunately, the $m=0$ modes contain very little power, so, rather than trying to distinguish between these types of quadratic QNMs, we simply do not include them in the search for QNMs. We note, though, that this can be resolved via a kind of post-processing, which we outline in App.~\ref{app:memory}. And, again, once the quadratic QNM ratios have been computed for all linear QNM combinations, this can be replaced.
	\end{enumerate}
	While the above points are the more-important of the less-important details, we encourage the interested reader to read the full documentation of the algorithm in the GitHub repository~\href{https://github.com/keefemitman/qnmfinder}{\textcolor{gray}{\faGithubSquare}} to see other minor details.
	
	\figGaussianTest
	
	\figNonGaussianTest
	
	We find that the speed of the algorithm is dependent on the fitting start time step size $\Delta t_{0}$ as well as the type of binary considered. For $\Delta t_{0}=1M$ on a non-precessing binary black hole system, it takes $\sim10$ minutes to extract  $\sim 40$ QNMs. However, for smaller $\Delta t_{0}$ or for highly precessing systems in which many QNMs can be loudly excited, the algorithm can sometimes take up to $\sim1$ hour. We again encourage the interested reader to look 
	at the GitHub repository~\href{https://github.com/keefemitman/qnmfinder}{\textcolor{gray}{\faGithubSquare}} for an example Jupyter notebook showing the algorithm's typical performance.
	
	\subsection{Analytic tests}
	\label{sec:analytictests}
	
	Often, these types of greedy algorithms are validated by seeing how well they can recover some set of QNMs from a waveform for which the QNMs \emph{are} known a priori. In line with these previous works, in Fig.~\ref{fig:gaussian_test} we perform a similar test, in which a waveform is constructed from a set of QNMs with some Gaussian noise added on top. Specifically, we imagine modeling a precessing system (for which the amplitudes of mirror QNMs, i.e., the two QNMs with $(\ell,m,n,p)$ and $(\ell,-m,n,p)$, need not be equal to each other). The 44 arbitrary QNMs we consider are
	\begin{align}
		\label{eq:testQNMs}
		&\{(2-3,0\leq m\leq\ell,0,\pm)\} + \{(4,0\leq m\leq4,0,+)\}\nonumber\\
		&+ \{(2,2,1-3,+)\} + \{(2,1,1,+)\}\nonumber\\
		&+ \{[(2,2,0,+),(2,2,0,+)]\} + (m\rightarrow-m)
	\end{align}
	to probe how the algorithm behaves across a wide spread of different harmonics, progrades/retrogrades, overtones, and quadratic QNMs. To obtain reasonable amplitudes for the test (as opposed to just random amplitudes), we fit these QNMs to a mass ratio $q=4$, precessing simulation (see Sec.~\ref{sec:NRwaveforms} for more details). Furthermore, we add white noise with $\sigma^{\mathrm{noise}}=10^{-5}$. The exact amplitudes as a function of time, as well as the noise, can be seen in the left-most panel of Fig.~\ref{fig:gaussian_test}. In the middle panel, we show the QNM model built by the algorithm, fit across various start times $t_{0}$ with the region of stability highlighted by thick lines. We take the tolerances to be $\Delta A^{\mathrm{min}}=158.0$ and $\sigma^{\mathrm{max}}=0.2$ (see Eqs.~\eqref{eq:timerequirement} and~\eqref{eq:variationrequirement}).\footnote{Note that we show these tolerances because these are close to the tolerances that we pick for analyzing NR simulations based on the results presented in Fig.~\ref{fig:non_gaussian_test}. The value of $\Delta A^{\mathrm{min}}=158.0$ simply comes from one of the middle bins when logarithmically binning the interval from $10$ to $10^{3}$ (see Fig.~\ref{fig:non_gaussian_test}).} Every QNM in the test waveform is recovered by the algorithm, and the amplitudes are consistent up to noise fluctuations. In the right-most panel, we plot the residual error as a function of start time, where we define this error as
	\begin{align}
		\label{eq:residualerror}
		\Delta(h,h_{\mathrm{QNM}},t_{1},t_{f})=\frac{\sqrt{\int_{t_{1}}^{t_{f}}\int_{S^{2}}|h-h_{\mathrm{QNM}}|^{2}\,d\Omega\,dt}}{(t_{f}-t_{1})},
	\end{align}
	with $t_{f}$ the end of the waveform. As can be seen, across all times the error is consistent with the magnitude of the injected noise. Thus, the algorithm can clearly handle this simple toy problem with Gaussian white noise.
	
	However, NR simulations do not have Gaussian noise. In fact, their ``noise'' is not the noise that experimentalists are familiar with, but is instead some systematic related to the truncation error of the code (or, e.g., unphysical boundary conditions~\cite{Ma:2024hzq}). The truncation error, though, is really a combination of spacial discretization errors, which stem from using spectral methods and adaptive mesh refinement~\cite{Berger:1984zza}, and temporal discretization errors, which stem from time steppers. So, to test the algorithm against a perhaps more realistic scenario, in Fig.~\ref{fig:non_gaussian_test} we study how the algorithm responds to a test waveform whose noise is more emblematic of NR truncation error. Specifically, we make the straightforward connection that the truncation error arising from the spatial discretization in an NR simulation is similar to truncating the series in Eq.~(51) of Ref.~\cite{Cook:2014cta} at $\ell'=\ell_{\mathrm{max}}$ (this simply corresponds to truncating the matrix used in the spectral eigenvalue approach of computing QNM frequencies). Such a cut introduces errors in the QNM frequencies that are larger for higher-frequency QNMs. To push this test to more of an extreme, we take $\ell_{\mathrm{max}}=4$ instead of the commonly used value of $\ell_{\mathrm{max}}=20$. This results in relative errors in the QNM frequencies on the order of $\mathcal{O}(0.1)$ for the lower-frequency QNMs and $\mathcal{O}(1)$ for the higher-frequency. The exact ``noise'' we inject is then each input QNM's amplitude scaled by $\sigma^{\mathrm{noise}}$ times the damped exponential matching this incorrect QNM frequency, i.e.,
	\begin{align}
		h_{\mathrm{noise}}=\sigma^{\mathrm{noise}}\sum\limits_{k}e^{i\tilde{\phi}_{k}}C_{k}e^{-i\tilde{\omega}_{k}t},
	\end{align}
	where $k$ ranges over the QNMs in Eq.~\eqref{eq:testQNMs}, $\tilde{\phi}_{k}$ is a random phase for each QNM, and $\tilde{\omega}_{k}$ is the incorrect QNM frequency resulting from the matrix truncation in the spectral eigenvalue frequency calculation. An example can be seen in the left-most plot of Fig.~\ref{fig:non_gaussian_test}, for which the truncation was done using the python package \texttt{qnm}~\cite{Stein:2019mop}.
	
	\figNRExample
	
	Like the Gaussian noise test, we again find that for fairly large non-QNM injections, the algorithm is able to successfully recover every QNM in the analytic waveform with amplitudes comparable to the true values. This is highlighted through the middle and right-most panels in the top of Fig.~\ref{fig:non_gaussian_test}. In the bottom of Fig.~\ref{fig:non_gaussian_test} we show the success rate of the algorithm for various tolerance choices for $\sigma^{\mathrm{max}}$ (see Eq.~\eqref{eq:variationrequirement}) and $\Delta A^{\mathrm{min}}$ (see Eq.~\eqref{eq:timerequirement}), with each colored square being a different pair of tolerances. Blue squares correspond to every QNM being recovered, light red correspond to some of the QNMs being missed, and dark red correspond to spurious QNMs, i.e., QNMs not in the input waveform, being found by the algorithm. Obviously the latter is the situation that we want to avoid the most, though finding every QNM that is present in the waveform is also key to constructing accurate models. We find that these spurious QNMs typically correspond to quadratic QNMs whose frequency is close to overtones. Therefore, based on these results for extractions with NR simulations, we choose to use the modest tolerances $\sigma^{\mathrm{max}}=0.2$ and $\Delta A^{\mathrm{min}}=100$. This equates to requiring that the complex QNM amplitude not vary by more than $20\%$ over a time interval in which it decays by two orders of magnitude. For fundamental QNMs this interval is roughly $50M$, while for $n=1$ and $n=2$ overtones it is $\sim20M$ and $\sim10M$. To see how these tolerances impact our ability to extract QNMs from simulations that have been previously analyzed in the literature, like SXS:BBH:2423 in Ref.~\cite{Giesler:2024hcr}, see App.~\ref{app:reproduce}.
	
	\section{Results}
	\label{sec:results}
	
	\figParameterSpaceQNMs
	
	With our $\sigma^{\mathrm{max}}$ (see Eq.~\eqref{eq:variationrequirement}) and $\Delta A^{\mathrm{min}}$ (see Eq.~\eqref{eq:timerequirement}) tolerances set to be $\sigma^{\mathrm{max}}=0.2$ and $\Delta A^{\mathrm{min}}=100$, as motivated by the tests presented in Sec.~\ref{sec:analytictests}, we now turn to extracting meaningful information about the excitation of QNMs from NR simulations using the algorithm. First, we study two astrophysically-relevant BBH simulations, and then we study a wide range of BBH mergers across the quasi-circular, non-precessing part of parameter space.
	
	\subsection{Numerical relativity waveforms}
	\label{sec:NRwaveforms}
	
	Before fitting numerical relativity waveforms with what black hole perturbation theory predicts, it is crucial to ensure that the NR simulation is in the canonical frame that is used when doing black hole perturbation theory. Every waveform, whether from an observation, simulation, or theory, effectively lives at future null infinity $\mathcal{I}^{+}$---the conformal completion of some physical spacetime, i.e., the final destination of outgoing radiation~\cite{Bondi1960,Sachs1961,Bondi:1962px,Sachs:1962wk,Sachs1962PR}. Consequently, one can always transform the waveform by whatever symmetries exist at $\mathcal{I}^{+}$ without changing the underlying physics. These symmetries are elements of the BMS group, which extends the Poincaré group to include an infinite number of direction-dependent translations called supertranslations~\cite{Bondi1960,Sachs1961,Bondi:1962px,Sachs:1962wk,Sachs1962PR,Mitman:2024uss}.
	
	So, because every NR waveform is effectively in some arbitrary frame (resulting from however the initial data is constructed and whatever gauge choice is made for the evolution), before comparing to perturbation theory's predictions it is important to map the simulation to the canonical frame of an isolated black hole. This is often referred to as the \emph{superrest frame}~\cite{OMMoreschi_1986,Moreschi:1988pc,Moreschi:1998mw,Dain:2000lij,Mitman:2021xkq,MaganaZertuche:2021syq,Mitman:2022kwt,Mitman:2024uss}, and is simply an extension of the usual center-of-mass frame to be shear-free: having the strain be zero as $t\rightarrow\infty$. Therefore, before analyzing an NR waveform, we map the data to the superrest frame of the final black hole by minimizing the appropriate BMS charges using the python package \texttt{scri}~\cite{boyle_2024_14531184,Boyle:2013nka,Boyle:2014ioa,Boyle2016}. Furthermore, to avoid errors induced by residual supertranslations stemming from the inaccuracy of the NR data, rather than fitting the strain we fit the news, i.e., the first time derivative of the strain, since the news always decays to zero regardless of what frame it is in~\cite{Boyle2016}. We note, though, that the amplitude of the QNMs in the strain-domain can easily be recovered from news-domain fits via\footnote{Note that in Eq.~\eqref{eq:integratedamplitudes}, no integration constant in needed since the system is in a frame which is shear-free as $t\rightarrow+\infty$.}
	\begin{align}
		\label{eq:integratedamplitudes}
		C_{\mathrm{QNM}}^{\mathrm{strain}}&=\left(\int C_{\mathrm{QNM}}^{\mathrm{news}}e^{-i\omega_{\mathrm{QNM}}t}\,dt\right)/e^{-i\omega_{\mathrm{QNM}}t}\nonumber\\
		&=C_{\mathrm{QNM}}^{\mathrm{news}}/(-i\omega_{\mathrm{QNM}}).
	\end{align}
	
	The remnant mass and spin of each simulation are computed using the appropriate BMS charges, following the same procedure outlined in Ref.~\cite{Iozzo:2021vnq}.
	
	Furthermore, to avoid higher mode content that may be poorly resolved, before fitting NR waveforms we truncate them so that their maximum $\ell$ is $\ell_{\mathrm{max}}=6$ unless stated otherwise in the text, like in App.~\ref{app:reproduce}.
	
	All of the simulations considered in this work were created using the SXS Collaboration's code \texttt{SpEC}~\cite{Boyle:2019kee} and their waveforms and Weyl scalars were extracted to $\mathcal{I}^{+}$ using the public \text{SpECTRE} code's CCE module~\cite{Moxon:2020gha,Moxon:2021gbv,spectrecode}. Their data will be made publicly available in the upcoming SXS Collaboration's CCE catalog.
	
	\subsection{Numerical relativity fits}
	\label{sec:NRfits}
	
	We begin by studying two BBH merger simulations that are astrophysically relevant: a mass ratio $q=1$ system with dimensionless spins in the direction of the orbital angular momentum $(\hat{z})$ with $\chi_{1}^{(z)}=-\chi_{2}^{(z)}=0.6$ and a mass ratio $q=4$ precessing system. Both of these systems are quasi-circular binaries and their data is publicly available in the EXT-CCE Waveform Database of the SXS Catalog~\cite{Boyle:2019kee}. In Fig.~\ref{fig:nr_example} we show the QNMs that are found by the algorithm when it is run on the news waveform of each simulation in the superrest frame of the remnant black hole. The top panels show the model that is recovered, with thin lines being the fits across all times and thick lines being the first $5M$ window over which the QNM's amplitude agrees with the amplitude extracted over $\Delta t_{\mathrm{QNM}}$ within $10\%$. The legend shows the QNMs in the order in which they are found. Only QNMs with $m\geq0$ are plotted, but the mirror QNMs with $m<0$ and the same $p$ value are also found. The bottom panels show errors over the two-sphere, like the residuals and the mismatch $\mathcal{M}$, which we define as
	\begin{align}
		\label{eq:mismatch}
		\mathcal{M} = 1 - \frac{\langle h_{1},h_{2}\rangle}{\sqrt{\langle h_{1}, h_{1}\rangle\langle h_{2}, h_{2}\rangle}},
	\end{align}
	where
	\begin{align}
		\langle h_{1},h_{2}\rangle=\mathrm{Re}\left[\int_{t_{1}}^{t_{f}}\int_{S^{2}} h_{1}^{*}h_{2}\,d\Omega dt\right].
	\end{align}
	Note that this is different from the mismatch usually presented in the literature, as here we are comparing the waveforms over the entire two-sphere and not performing any time or phase alignment before computing the error.
	
	In the top panel, we include these thick lines to help provide a rough proxy for the onset of the QNM's stability, which could be of interest to certain LVK data analyses. These regimes may be interpreted as the onset of stability, defined by the mean amplitude over a $5M$ window having a relative error of $10\%$ with the most stable amplitude, with respect to the QNM model found by the algorithm.\footnote{This latter point is a key caveat, as the time of this regime can vary by several $M$ depending on the QNM model fit to the data. In particular, if the near-peak physics is modeled better, e.g., with the inclusion of more overtones, then this regime occurs at earlier times; see, e.g., Fig.~6 of Ref.~\cite{Giesler:2024hcr}.} For extracting the amplitudes, however, we always use the window of size $\Delta t_{\mathrm{QNM}}$ over which $\sigma_{\mathrm{QNM}}$ is minimized, to ensure that we are sufficiently deep in the ringdown stability regime, rather than somewhere near the onset. These amplitudes measured over $\Delta t_{\mathrm{QNM}}$ are the ones used to compute the full QNM model, which is compared to the NR waveform in the bottom panels.
	
	As can be seen, for the $q=1$, $\chi_{1,2}^{(z)}=\pm0.6$ system, the algorithm manages to find QNMs in the dominant even $m$ modes as well as the odd $m$ modes, as is expected for such an asymmetric system. Furthermore, the algorithm also finds the $n=1$ and $n=2$ $(2,2,n,+)$ overtones, the $n=1$ $(2,1,n,+)$ overtone, and the $(2,2,0,+)\times(2,2,0,+)$ quadratic QNM. In the magenta curve in the lower panel, one sees that the mismatch begins at $\mathcal{O}(1)$ near the time of peak strain and quickly decays to $\mathcal{O}(10^{-2})$ in $\sim10 M$. Consequently, not only is the algorithm able to find many stable QNMs, but it also does so in a manner of modeling the waveform that is accurate enough for detector analysis purposes, provided one waits a few $M$ past the time of peak strain over the two-sphere (see Eq.~\eqref{eq:peakstrain}).
	
	In the right panel of Fig.~\ref{fig:nr_example}, we show results for the $q=4$, precessing system. As expected~\cite{Finch:2021iip,Zhu:2023fnf}, many more QNMs are found with power much more on par with the usual $(2,2,0,+)$ QNM. While no $n=2$ overtones or quadratic QNMs are found, the mismatch is on par with the non-precessing system, and thus points to the fact that QNM excitation, both fundamentals, overtones, and quadratics, is highly dependent on the binary black holes' progenitor parameters. In Sec.~\ref{sec:QCNPsystems} we expand on this.
	
	Lastly, we also want to note that the the algorithm also finds the identical set of QNMs for each system for lower resolution versions of the simulations (as shown through the dashed yellow curves in the bottom panels). Therefore, our confidence that the algorithm is extracting meaningful QNMs is only furthered.
	
	\subsection{Quasi-circular, non-precessing system fits}
	\label{sec:QCNPsystems}
	
	\figParameterSpaceMismatches
	
	With the algorithm's result on two NR simulations matching the QNMs that we would naively expect from previous works~\cite{Dhani:2020nik,Dhani:2021vac,Li:2021wgz,Finch:2021iip,Ma:2022wpv,Zhu:2023fnf,Mitman:2022qdl,Cheung:2022rbm,Giesler:2024hcr}, we now switch to examining the algorithm's behavior over a set $\sim400$ quasi-circular, non-precessing simulations with $q\leq8$ and $|\chi_{1,2}^{(z)}|<0.8$. In Fig.~\ref{fig:parameter_space_QNMs} we show the excitation of various QNMs that the algorithm finds as a function of
	\begin{align}
		\label{eq:chieff}
		\chi_{\mathrm{eff}}=\frac{q\chi_{1}^{(z)}+\chi_{2}^{(z)}}{1+q}
	\end{align}
	and  
	\begin{align}
		\label{eq:chia}
		\chi_{\mathrm{a}}=\frac{q\chi_{1}^{(z)}-\chi_{2}^{(z)}}{1+q},
	\end{align}
	with marker size representing the mass ratio. Each of the six panels toward the top of the figure correspond to a different QNM amplitude or a QNM amplitude ratio. Color indicates amplitude magnitude, with red markers indicating a simulation for which the QNM was not found. As can be seen, the algorithm does particularly well at modeling the expected physics, e.g., the $(2,1,0,+)$ QNM is never found in $q=1$ simulations with $\chi_{a}=0$ (which is what we expect from symmetry~\cite{Kamaretsos:2011um,Kamaretsos:2012bs}) and the $(2,2,0,-)$ QNM is predominantly found in simulations that have $\chi_{\mathrm{eff}},\chi_{a}\ll0$~\cite{Dhani:2020nik,Dhani:2021vac,Li:2021wgz,Ma:2022wpv}. Furthermore, panels d) - f) of Fig.~\ref{fig:parameter_space_QNMs} also show that the algorithm is able to recover $n\leq3$ overtones, albeit for certain progenitor parameters (see App.~\ref{app:reproduce} for an example of finding the $n=4$ overtone).
	
	The bottom panel of Fig.~\ref{fig:parameter_space_QNMs} shows the start time of the first $5M$ window in which the mean amplitude matches the most stable amplitude within 10\%---a rough proxy for the onset of the QNM's stability---for the QNMs that are found at least $20$ times across the $\sim400$ simulations by the algorithm. Put differently, each QNM's violin plot is the union (over the simulations) of the left-most time of the thick line in Fig.~\ref{fig:nr_example} for the QNM. Mathematically, this onset time $t_{\mathrm{QNM}}^{\mathrm{onset}}$ is computed via
	\begin{align}
		\label{eq:tonset}
		&t_{\mathrm{QNM}}^{\mathrm{onset}}\equiv\nonumber\\
		&\mathrm{min}_{t}\left(\left\{\left[t,t+5M\right]\Bigg|\frac{|C_{\mathrm{QNM}}^{[t,t+5M]}-C_{\mathrm{QNM}}^{\Delta t_{\mathrm{QNM}}}|}{|C_{\mathrm{QNM}}^{\Delta t_{\mathrm{QNM}}}|}<0.1\right\}\right).
	\end{align}
	As can be seen, the $(2,2,0,+)$ matches its most stable amplitude over a $5M$ window very early on---within a few $M$ of the peak; meanwhile, the higher harmonics, overtones, and quadratic QNMs, tend to require more time after the peak of the strain for this stability criterion to be met. In an arbitrary QNM column, we find that the simulations that satisfy this criterion later tend to be those with a more highly spinning remnant, in agreement with Ref.~\cite{Giesler:2024hcr}. Furthermore, we find that the time at which this criterion is satisfied is highly dependent on the QNM model that the algorithm ends up constructing for a given simulation. In particular, if more faster-decaying QNMs are recovered, the criterion tends to be satisfied earlier in time, presumably because the near-peak physics is better modeled in this situation. In the future, it would be interesting to explore the onset of stability in much more detail, as this is key for GW event data analyses.
	
	The horizontal axis of the bottom panel of Fig.~\ref{fig:parameter_space_QNMs} also shows that there is a very rich spectrum of stable QNMs. A few noteworthy ones are perhaps the $(2,0-2,n,\pm)$, $(3,2-3,n,+)$, and $(4,4,n,+)$ overtones, many of the retrograde QNMs, the previously found quadratic QNMs $(2,2,0,+)\times(2,2,0,+)$ and $(2,2,0,+)\times(3,3,0,+)$, and the new $(2,2,0,+)\times(4,4,0,+)$, $(2,1,0,+)\times(2,2,0,+)$, and $(2,2,0,+)\times(2,2,1,+)$ quadratic QNMs. While not shown on the horizontal axis, as stated earlier we find overtones up to $n=4$ in the $(2,2)$ mode, though only in a few simulations, one of which is shown in App.~\ref{app:reproduce}.
	
	To provide more meaning to these extracted QNMs and their variations in the context of waveform modeling, in the left panel of Fig.~\ref{fig:parameter_space_mismatches} we show the mismatch between the QNM model that the algorithm is able to construct and the NR waveform at the post-peak time $t_{\mathrm{peak}}^{\mathrm{strain}}+10M$. Evidently, the mismatch seems to be fairly uniform across parameter space, with some slight bias for large $\chi_{\mathrm{eff}}$ and $|\chi_{a}|$ systems, which is reassuring that the majority of the progenitor-dependent physics is being captured.
	
	The right panel of Fig.~\ref{fig:parameter_space_mismatches} shows the mismatch $\mathcal{M}$ (black) and median residual (blue) between the model and the NR data across progressively later times. The gray region shows the minimum/maximum mismatches, while the thick line shows the median. With the magenta curve, we show the mismatch from the $q=4$, precessing system in the right panel of Fig.~\ref{fig:nr_example} to show that the algorithm also performs similarly well for precessing systems as it does for this large number of non-precessing systems.
	
	The more interesting curve, however, is the blue curve. This median residual---which for each time is the median of the residual between the QNM model and the NR data across all simulations---corresponds to whatever the dominant unmodeled contribution to the NR waveform is at the corresponding time $t_{1}$. As a reference, in the dotted gray lines we plot the typical decay times of the $n=0,1$ and $2$ overtones as well as the quadratic, cubic, and quartic QNMs. As illustrated, at early times the error seems to fall off with respect to the $n=2$ overtone, which suggests that the limiting factor in our models' accuracy may be the lack of more $n=2$ QNMs across our catalog of simulations (see Refs.~\cite{Takahashi:2023tkb,Gao:2025zvl} for in depth studies of how residuals can point to missing QNM content). However, to confirm this without simply lowering our $\sigma^{\mathrm{max}}$ tolerance, one would presumably need to incorporate better fitting techniques, like the mode dropping used in Ref.~\cite{Giesler:2024hcr}, or even more physics, like an automated inclusion of quadratic QNMs or a model of the prompt response, which we reserve for future work once more of this physics is understood.
	
	\subsection{Probing the remnant perturbation's structure}
	\label{sec:EFanalysis}
	
	\figOvertoneExcitation
	
	Finally, with an understanding of the algorithm's ability to extract QNMs across quasi-circular, non-precessing parameter space as well as its accuracy and limitations, we now turn to studying what some of the QNMs that we have extracted may tell us about the structure of remnant perturbations following a BBH merger. One such study is probing what the metric perturbation to the remnant black hole looks like, as it has been suggested that it is not as complicated as imagined~\cite{Kamaretsos:2012bs,Borhanian:2019kxt,Oshita:2021iyn,Zhu:2023fnf,Zhu:2024rej,Giesler:2024hcr}.
	
	\subsubsection{Overtone excitation factor}
	
	According to perturbation theory, the Kerr geometry tends to excite certain QNMs more than others through their \emph{excitation factors}~\cite{Leaver:1986gd}. More explicitly, any QNM's complex amplitude can be written as the product of a source $I_{\mathrm{QNM}}$ and its excitation factor $E_{\mathrm{QNM}}$, i.e.,
	\begin{align}
		\label{eq:excitationfactor}
		C_{\mathrm{QNM}}=I_{\mathrm{QNM}}E_{\mathrm{QNM}},
	\end{align}
	where $I_{\mathrm{QNM}}$ is a spatial integral over the source and $E_{\mathrm{QNM}}$ is a factor in the time-domain Green's function (see, e.g., Refs.~\cite{Berti:2006wq,Zhang:2013ksa}). Interestingly, Refs.~\cite{Giesler:2019uxc,Oshita:2021iyn} found that the norms of the complex amplitudes of the $(2,2,n,+)$ overtones extracted from NR simulations seem to track the excitation factors as a function of $n$ closely up to $n=7$. This result implies that the amplitudes of prograde $(2,2)$ QNMs at late times behave as if they had a source term at earlier times which was independent of the index $n$, i.e., $I_{(\ell,m,n,p)}\sim I_{(\ell,m,0,p)}$. Furthermore, in Ref.~\cite{Giesler:2024hcr} even more evidence related to this observation was presented, albeit for only two simulations. Ref.~\cite{Cheung:2023vki} also showed similar findings, although restricted to $n=1$ and affected by some other limitations that we will address herein. Confirming that such trends hold over the entirety of binary parameter space would have implications for the structure of remnant perturbations in BBHs. So, in Fig.~\ref{fig:overtone_excitation}, we revisit this question using the QNM amplitudes that have been extracted by our algorithm.
	
	\fitExcitationParameters
	
	Because of our stability criterion, our QNM amplitude fits are obtained $\mathcal{O}(10M)$ after $t_{\mathrm{peak}}^{\mathrm{strain}}$. If we assume that: the source function $I_{\mathrm{QNM}}$ in Eq.~\eqref{eq:excitationfactor} is independent of the index $n$ for the $(2,2,n,+)$ QNMs; the time-dependent initial excitation of individual QNMs~\cite{Chavda:2024awq} can be ignored when we extrapolate our fitted amplitudes to earlier times; and the fundamental and overtone QNMs sharing a given set of $(2,2,\cdot,+)$ indices are all excited simultaneously; then we should be able to extrapolate the amplitudes of the different $(2,2,n,+)$ QNMs from when they are fit to a single time where the norms of their amplitudes have the same dependence on $n$ as the excitation factors $E_{\mathrm{QNM}}$, up to a multiplicative complex constant.\footnote{Ref.~\cite{Oshita:2024wgt} introduced a convention in which a time ambiguity is included in the excitation factors. This ambiguity, however, can always be absorbed into the source function $I_{\mathrm{QNM}}$, making the excitation factors $E_{\mathrm{QNM}}$ instead be time-independent numbers. For conceptual clarity, we will treat the excitation factors $E_{\mathrm{QNM}}$ as time-independent, in line with the conventions of Refs.~\cite{Berti:2006wq,Zhang:2013ksa}.} This means we expect to be able to find some $t_{\mathrm{pert.}}^{(2,2)}$ such that
	\begin{align}
		\label{eq:normalizedamplitudes}
		E_{(2,2,n,+)} \sim \frac{1}{I_{(2,2)}}e^{-i\omega_{(2,2,n,+)}t_{\mathrm{pert.}}^{(2,2)}}C_{(2,2,n,+)}
	\end{align}
	for all $n$, with $I_{(2,2)}$ the multiplicative complex constant. Note that because the amplitudes $C_{(2,2,n,+)}$ are defined at the time of peak strain, $t_{\mathrm{pert.}}^{(2,2)}$ naturally represents differences with respect to that time. Such a time $t_{\mathrm{pert.}}^{(2,2)}$ (and complex constant $I_{(2,2)}$) would obey the equations:
	\begin{subequations}
	\label{eq:EFparameters}
	\begin{align}
		t_{\mathrm{pert.}}^{(2,2)}=\frac{\ln\left(\left|\frac{E_{(2,2,0,+)}C_{(2,2,1,+)}}{E_{(2,2,1,+)}C_{(2,2,0,+)}}\right|\right)}{\mathrm{Im}[\omega_{(2,2,0,+)}-\omega_{(2,2,1,+)}]};\\
		I_{(2,2)}=\frac{C_{(2,2,0,+)}}{E_{(2,2,0,+)}}e^{-i\omega_{(2,2,0,+)}t_{\mathrm{pert.}}^{(2,2)}}.
	\end{align}
	\end{subequations}
	Note that through Eqs.~\eqref{eq:EFparameters}, the amplitudes of the $n=0$ and $n=1$ QNMs and the phase of the $n=0$ QNM are made to match their corresponding excitation factors. Also, the time $t_{\mathrm{pert.}}^{(2,2)}$ and constant $I_{(2,2)}$ need not (and in fact should not) be independent of the binary parameters. Thus, for each simulation we solve for $t_{\mathrm{pert.}}^{(2,2)}$ and $I_{(2,2)}$ independently. This is more in line with the analysis that was performed in Ref.~\cite{Giesler:2024hcr}, and differentiates our result from that presented in Ref.~\cite{Cheung:2023vki}.
	
	With these free parameters fixed, we can now proceed to examine the results of Fig.~\ref{fig:overtone_excitation}. In the left panel of Fig.~\ref{fig:overtone_excitation} we show every $(2,2,0-3,+)$ QNM amplitude found by the algorithm across the $\sim400$ simulations that we examined. Data points are the extracted amplitudes, with error bars corresponding to the standard deviation of the amplitude over its most stable window, while solid lines are the Teukolsky excitation factors from Refs.~\cite{Zhang:2013ksa,EFs1,EFs2}.\footnote{While the quasi-normal mode excitation factors for both the Teukolsky and Sasaki-Nakamura equations have been reported in the literature, it is incorrect to use the Sasaki-Nakamura formalism's excitation factors when comparing to NR simulations, unless they have been converted to match the convention of the Teukolsky formalism (see Refs.~\cite{Hughes:2000pf,Sasaki:1981sx,Sasaki:1981kj,Lo:2023fvv} and Eq. (53) of Ref.~\cite{Zhang:2013ksa}). Ref.~\cite{Cheung:2023vki} seems to use the Sasaki-Nakamura excitation factors, which is also why their results are slightly at odds with ours.}$\phantom{}^{,}$\footnote{We also scale the excitation factors in Refs.~\cite{EFs1,EFs2} by a factor of $1/(-i\omega_{\mathrm{QNM}})^{2}$ so that they match the strain instead of $\Psi_{4}$.}$\phantom{}^{,}$\footnote{In Ref.~\cite{Motohashi:2024fwt} it was pointed out that the excitation factors computed in Refs.~\cite{Zhang:2013ksa,EFs1,EFs2} are missing a factor of $e^{-i\omega(1-\sqrt{1-a^{2}})}$. Also, from private conversations with Aaron Zimmerman, there may be additional phase corrections that are missing. Even so, such corrections would not change our overall results except to induce mass and spin dependent changes to the values in Fig.~\ref{fig:excitation_parameters}.} As constrained by Eqs.~\eqref{eq:EFparameters}, the $n=0$ and $n=1$ QNM amplitudes and the $n=0$ QNM phase exactly match their excitation factors. The more interesting points are the unconstrained $n=2$ and $n=3$ QNMs. If the source term $I_{(2,2)}$ was some arbitrary function of $n$, then you would expect no relationship between the orange/magenta data and their same-colored corresponding excitation factors. However, as can be seen, the $n=2$ QNMs clearly track the excitation factors, as do the $n=3$ QNMs, albeit slightly less so. Therefore, this provides evidence that, at least for the $n\leq2$ and maybe even for the $n\leq3$ $(2,2,n,+)$ overtones, the perturbation to the remnant in quasi-circular, non-precessing BBH coalescences may be independent of the overtone index $n$. This provides much broader support for the previous observations made in Refs.~\cite{Giesler:2019uxc,Oshita:2021iyn,Cheung:2023vki,Giesler:2024hcr} that are related to this finding.
	
	Interestingly, in the right panel of Fig.~\ref{fig:overtone_excitation} where we show the phases of the complex QNM amplitudes and their excitation factors, while we see the expected agreement for $n=0$, there seems to be a trend of progressively larger disagreements with the excitation factors as $n$ increases. While it is not particularly surprising that the source may have some nontrivial phase dependence, this trend may also simply be unphysical. There are some hints of a similar trend in the amplitudes (see the $n=3$ points in the right panel of Fig.~\ref{fig:overtone_excitation}), but in Ref.~\cite{Giesler:2024hcr}, no such amplitude disagreement was found for this comparison. Consequently, since the overtones are highly correlated with one another, it could be that we are seeing this disagreement because not as many overtones are found by our algorithm as were included in the fits of Ref.~\cite{Giesler:2024hcr}. More work is required to fully understand if this result is because of the QNMs included in our fits, or really because of some structure in the remnant perturbation.

	We also find a very similar result for the $(3,3,n,+)$ QNMs, as shown in the bottom panel of Fig.~\ref{fig:overtone_excitation}, although the agreement is not as strong as for the $(2,2,n,+)$ QNMs.
	
	\paragraph{Interpretation}

	Our findings in this section may provide some insightful information about the structure of the remnant perturbations sourced in BBH mergers. As stated above, the agreement we found between the excitation factors and the $(2,2,n,+)$ overtones at time $t_{\mathrm{pert.}}^{(2,2)}$ was predicated on assumptions: one of which was assuming that we could safely ignore the time-dependent ring-up of the QNMs. One explanation as for why this assumption may be valid could be the following: the radial extent of the $(2,2,n,+)$ perturbation and/or its timescale of interaction with the peak of the scattering potential may be small, and in particular of order much smaller than the width of the scattering potential (i.e., much smaller than the radius of the remnant black hole). Such a condition could possibly cause the QNMs to be quickly excited before decaying, since the time-dependent ring-up of the QNMs has a timescale that is proportional to the radial extent of the remnant perturbation~\cite{Chavda:2024awq}.

	Furthermore, if one considers the time $t_{\mathrm{pert.}}^{(2,2)}$ we found to have some physical meaning, and if the perturbation has a relatively short-lived interaction with the remnant's scattering potential's peak, then one natural choice might be to interpret the time $t_{\mathrm{pert.}}^{(2,2)}$ as the time when the bulk of the perturbation is nearest to the potential's peak, i.e., a time reasonably close to when the $(2,2,n,+)$ QNMs are being excited~\cite{Oshita:2024wgt}. This would imply that at times close to the peak strain, content from perturbation theory---like the QNMs---may already be being sourced.

	The condition of a radially narrow perturbation could also limit the prompt response to be short lived (the prompt response might be expected to have a timescale on the order of the radial scale of the perturbation itself; see, e.g., Refs.~\cite{Andersson:1996cm, Zenginoglu:2012xe,Chavda:2024awq}). Combined with the possibility of QNMs being present close to the peak strain time, this could provide context for why previous studies such as Ref.~\cite{Giesler:2024hcr} have been able to find that QNM fits are a good description of the waveform close in time to the peak of the strain in binary black hole coalescences.
	
	Last, we may be able to extract even more information about the ringdown perturbation by studying the values of the parameters in Eqs.~\eqref{eq:EFparameters} needed to make the $n=0$ and $n=1$ QNMs match their excitation factors. In Fig.~\ref{fig:excitation_parameters} we present the values of $|I_{(2,2)}|$ and $t_{\mathrm{pert.}}^{(2,2)}$ as a function of remnant spin and mass ratio. The value of $|I_{(2,2)}|$ tells us how the amplitude of the source varies, while $t_{\mathrm{pert.}}^{(2,2)}$ may tell us something about the time at which the $(2,2,n,+)$ QNMs start to be excited~\cite{Oshita:2024wgt}.

	As illustrated, the source term $|I_{(2,2)}|$ exhibits what one may naively expect: as the remnant spin increases and more power enters into the higher $\ell$ harmonics, $|I_{(2,2)}|$ decreases; and, as the mass ratio increases and more power enters into the, e.g., odd $m$ harmonics, $|I_{(2,2)}|$ decreases. As for $t_{\mathrm{pert.}}^{(2,2)}$, one readily finds that as the spin increases, so does $t_{\mathrm{pert.}}^{(2,2)}$. While a physical interpretation of this is not exactly obvious, it could simply be connected to the fact that the time scale of the ringdown for more highly spinning remnants is longer, since the damping times of the QNMs increase as the spin increases. But it could be interesting to try to probe this behavior further.
	
	While our interpretive arguments are by no means rigorous, we believe they may still provide insight into the nature of ringdown perturbations in BBH mergers. We reserve a more formal analysis of these features for future work. Finally, it is also possible that such trends in mass ratio and remnant spin are not physics, but rather some kind of systematic effect related to our QNM fitting. That being said, the left panel of Fig.~\ref{fig:parameter_space_mismatches} shows that the algorithm seems to perform nearly-identically (in terms of waveform mismatch) across binary parameter space, so we believe this to likely not be the case.
	
	\subsubsection{Quadratic excitation factor}
	
	\figQuadratic
	
	Finally, to also probe the perturbation for the quadratic QNMs, we now examine the excitation of the $\begin{pmatrix}2&2&0&+\\2&2&0&+\end{pmatrix}\rightarrow(4,4)$ quadratic QNM relative to its two parent QNMs as well as the $\begin{pmatrix}2&2&0&+\\3&3&0&+\end{pmatrix}\rightarrow(5,5)$ quadratic QNM. In the top panels of Fig.~\ref{fig:quadratics} we show the relative amplitudes of these QNMs as a function of spin, while in the bottom panels we show the phase differences. Additionally, we plot the prediction from second-order black hole perturbation theory, computed in Refs.~\cite{Ma:2024qcv,Khera:2024bjs} (see Eq.~(29) of Ref.~\cite{Khera:2024bjs}). As is illustrated, within the error bars of our data, the values extracted from our simulations and those predicted by theory are in very reasonable agreement, suggesting that there are no other contributions to this quadratic QNM besides what one expects from perturbation theory. It would be interesting to compare the other quadratic QNMs found with the predictions from perturbation theory, but these values have yet to be computed more generally so we also save this study for future work.
	
	\section{Discussion}
	\label{sec:discussion}
	
	Understanding the QNM amplitude spectra excited in binary black hole coalescences is crucial for performing ringdown-based tests of GR and parameter estimation on massive BBH merger observations. However, extracting QNM amplitudes from NR simulations---the closest thing we have to exact GR---has proven to be challenging over the years due to imperfect NR waveforms and issues with fitting techniques for an ill-conditioned problem. While algorithms for extracting stable QNMs have been presented in the past, none have been able to recover the faster-decaying QNMs that have been shown to be both present and contain a large fraction of power in various binary black hole coalescence simulations~\cite{Giesler:2024hcr}.
	
	In this study we developed an algorithm for extracting stable QNMs from arbitrary BBH coalescence simulations which is able to identify faster-decaying QNMs, like the high overtones and quadratic QNMs. The main idea behind the algorithm is to utilize a much more advanced fitting algorithm, \texttt{VarPro}, to identify various best-fitting frequencies at late times, match them to QNMs, verify their stability, and then work towards progressively earlier times to identify the faster-decaying content that has yet to be identified. In doing so, we find that we can not only capture a significant number of QNMs that reflect the expected physics of the corresponding NR simulations, but can also construct a QNM model that yields a mismatch with NR $\mathcal{M}\lesssim\mathcal{O}(10^{-2})$ $\sim10M$ after peak strain.
	
	After verifying the algorithm against a number of test waveforms (for which the QNM content is already known), we study what QNMs the algorithm manages to extract across $\sim400$ quasi-circular, non-precessing simulations. Not only do we recover previously known relationships between certain QNMs and progenitor parameters, such as retrograde QNMs being excited for negatively-spinning black holes, but we also demonstrate the stability of the $n\leq 4$ overtones in certain parts of parameter space, thereby furthering the NR-based evidence for the need to use overtones when analyzing BBH merger detections.
	
	Lastly, by studying the amplitudes of the QNMs in relation to their excitation factors, we uncovered further evidence to support some previous suggestions that the amplitudes of prograde $(2,2)$ QNMs at late times behave as if they had a source term at earlier times which was independent of the overtone index $n$. We speculate that this observation may also reveal information about the spatiotemporal structure of the perturbation. It may even suggest that perturbative content, like QNMs, may be present in the data near the time of peak strain, and furthermore that the prompt response may be short-lived in binary black hole coalescences.
	
	Our algorithm, \emph{qnmfinder}, is made publicly available at the GitHub repository~\href{https://github.com/keefemitman/qnmfinder}{\textcolor{gray}{\faGithubSquare}}.
	
	\subsection{Future Work}
	\label{sec:futurework}
	
	In the future, we plan on modeling the QNMs extracted with this algorithm using Gaussian processes, much like the works of Refs.~\cite{Pacilio:2024tdl,MaganaZertuche:2024ajz}. However, we intend to not only produce a model for quasi-circular, non-precessing parameter space, but also for precessing systems, as it has already been shown in Fig.~\ref{fig:parameter_space_mismatches} that the algorithm performs similarly-well when extracting various QNMs from precessing systems, as for non-precessing systems.
	
	Finally, it would also be interesting to use the QNM models that this algorithm constructs to examine the breakdown of constant-QNM models as one moves toward earlier times. As clearly pointed out in Ref.~\cite{Andersson:1996cm,Chavda:2024awq}, even from first-order perturbation theory it is not expected that QNM amplitudes will be exactly stable at early times. Therefore, it would be interesting to see if we can probe when the constant-QNM picture is appropriate and to what degree. After all, if more physics about the ringdown can be understood, then we can better model this stage of BBH coalescences and thus use even more of the high signal-to-noise ratio data in BBH merger detections to perform potentially revolutionary tests of GR.
	
	\section*{Acknowledgments}
	\label{sec:acknowledgements}
	The authors thank Will Farr and Max Isi for useful and insightful discussions, Sizheng Ma and Neev Khera for clarifications regarding quadratic QNM excitations and for sharing the excitation factor data used in Fig.~\ref{fig:quadratics}, and Leo Stein for suggesting the non-Gaussian noise test.
	
	K.M.\ is supported by NASA through the NASA Hubble Fellowship grant \#
	HST-HF2-51562.001-A awarded by the Space Telescope Science Institute,
	which is operated by the Association of Universities for Research in
	Astronomy, Incorporated, under NASA contract
	NAS5-26555. H.S. is supported by Yuri Levin's Simons Investigator Award 827103. This material is based upon work supported by the National Science Foundation under Grants No.~PHY-2309211, No.~PHY-2309231, No.~OAC-2209656 at Caltech, and No.~PHY-2407742, No.~PHY-2207342, and No.~OAC-2209655 at Cornell. Any opinions, findings and conclusions or recommendations expressed in this material are those of the author(s) and do not necessarily reflect the views of the National Science Foundation. This work was supported by the Sherman Fairchild Foundation at Caltech and Cornell.

	\section*{Contributions}
	K.M. developed the published version of the algorithm, wrote the majority of the manuscript, and created all of the figures, besides Figure 1.
	I. P. developed an earlier version of the algorithm, conducted initial analyses on the stability of QNMs with respect to white noise,  performed tests to assess the algorithm's capabilities and accuracy, and also created Figure 1.
	H.S. principally contributed conceptual design, analysis, and writing to Sec.~\ref{sec:EFanalysis} and its associated figures.
		
	\appendix
	
	\section{Modeling Memory}
	\label{app:memory}
	
	\figMemory
	
	As pointed out in Sec.~\ref{sec:codeoutline}, because the algorithm models modes with a large amount of unmodeled power using Eq.~\eqref{eq:unmodeledpower}, even though the overall amplitude of $m=0$ modes can be very large due to the memory effect, they do not receive much attention from the algorithm. Consequently, if one computes mismatches between model and NR strain waveforms, then they can obtain much worse values than if the news waveforms were used. 
	
	In this section we wish to briefly point out how this can be mitigated using knowledge of the memory's relation to the strain. In particular, because the memory contribution to the strain goes as Eq.~(17b) of Ref.~\cite{Mitman:2020bjf} (see also Refs.~\cite{Talbot:2018sgr,Mitman:2020pbt,Mitman:2024uss}), one can correct the QNM model to contain memory by computing the memory contribution and simply adding it. In Fig.~\ref{fig:memory}, we demonstrate this.
	
	In the top panel of Fig.~\ref{fig:memory} we show the $(2,0)$ mode of the strain for a $q=1$, $\chi_{1,2}^{(z)}=0.6$ system, i.e., a BBH which is known to produce a very large memory signal. In black we show the NR waveform, in blue we show the QNM model returned by the algorithm, and in green we plot the QNM model, corrected by Eq.~(17b) of Ref.~\cite{Mitman:2020bjf}. As is clear, and as pointed out by the residual plots in the bottom panel, performing this memory correction to the QNM model built by the algorithm significantly improves the modeling of the strain waveform. Therefore, whenever ringdown models are constructed using this algorithm, we strongly encourage the creators to perform this correction to better model the full strain waveform, if desired.
	
	\section{SXS:BBH:2423}
	\label{app:reproduce}
	
	\figGiesler
	
	To put our algorithm in the context of previous works, here we present how the algorithm behaves on the SXS:BBH:2423 waveform that was analyzed in Ref.~\cite{Giesler:2024hcr}. This simulation is a quasi-circular, non-precessing system that yields a very highly spinning remnant ($\chi_{f}\sim0.95$). The corresponding waveform, apart from being mapped to the superrest frame of the remnant black hole, also has the memory contribution to the waveform removed so that one can focus on extracting non-memory QNMs.
	
	If Fig.~\ref{fig:giesler} we show the QNMs found by the algorithm when run on the $(2,\pm2)$ harmonics of this waveform.\footnote{We find that the higher $\ell$ harmonics of this waveform have more noise and make it harder to extract QNMs from the $(2,\pm2)$ modes.} As can be seen in the legend, the algorithm is able to find the $(2,2,n\leq4,+)$ and $(3,2,0,+)$ QNMs as well as the $(2,-2,0,+)\times(4,4,0,+)$ quadratic QNM that was also found in Ref.~\cite{Giesler:2024hcr}. While Ref.~\cite{Giesler:2024hcr} managed to find many more overtones as well as some other quadratic QNMs, we find that at our current tolerance of $\sigma^{\mathrm{max}}=0.2$ and without incorporating mode dropping, we cannot recover the same QNMs as Ref.~\cite{Giesler:2024hcr}. Retaining modes in the fitting procedure after their amplitudes have decayed affects the stability of the fit for the modes that are still present, an example of the ill-conditioned nature of the fitting problem. We, however, found it difficult to implement an automated way of dropping QNMs once their amplitudes have sufficiently decayed. Consequently, we leave the idea of mode dropping, as was adopted in Ref.~\cite{Giesler:2024hcr}, for the topic of future work.
	
	\bibliography{refs}
	
\end{document}